\documentclass[11pt]{article}
\pdfoutput=1
\usepackage{soul}
\usepackage{jheppub}
\usepackage{amsfonts}
\usepackage{amsmath}
\allowdisplaybreaks[4]         
\usepackage{amssymb}   
\usepackage{euscript}       
\usepackage{xcolor}          
\usepackage{tensor}     
\usepackage{caption}
\usepackage{subcaption}   

\usepackage{tikz}
\usepackage{graphicx}
\usetikzlibrary{calc,fadings,decorations.pathreplacing}
\usepackage{verbatim}

\usepackage{url}
\DeclareMathOperator{\MyProd}{\scalebox{1.4}{$\mathrm{I\kern-0.2ex I}$}}
\tikzset{%
  >=latex, % option for nice arrows
  inner sep=0pt,%
  outer sep=2pt,%
  mark coordinate/.style={inner sep=0pt,outer sep=0pt,minimum size=3pt,
    fill=black,circle}%
}

\newcommand{\ba}{\begin{eqnarray}}
	\usepackage{braket}
	\newcommand{\ea}{\end{eqnarray}}

\newcommand{\beq}{\begin{equation}}
\newcommand{\beqa}{\begin{eqnarray}}

\newcommand{\beqar}{\begin{eqnarray*}}

\newcommand{\be}{\begin{equation}}
\newcommand{\ads}{a_d^*}

\newcommand\wO{{\sigma}}

\newcommand\wT{{\tau}}

\newcommand\indfd{{\mathcal{I}_{4d}}}

\usepackage{comment}
\newcommand\fft[2]{\frac{#1}{#2}}

\def\i{\mathrm{i}}

\def\beqa{\begin{eqnarray}\displaystyle}
\def\eeqa{\end{eqnarray}}
\def\be{\begin{equation}}
\def\ee{\end{equation}}

\def\bse{\begin{subequations}}
\def\ese{\end{subequations}}

\newcommand{\bem}{\begin{pmatrix}}
\newcommand{\eem}{\end{pmatrix}}

\def\+{\, + \,}

\def\bar{\overline}
\def\ads2{AdS$_2$}
\def\ss2{S$^2$}

\def\rt2{\sqrt{2}}

\renewcommand{\Im}{\mbox{Im}}
\renewcommand{\Re}{\mbox{Re}}

\title{\boldmath Thermodynamics of black holes with probe D-branes}

\author[a]{Alejandro Cabo-Bizet}

\author[b]{Marina David}

\author[c]{Alfredo Gonz\'alez Lezcano}

\affiliation[a]{Università del Salento, Dipartimento di Matematica e Fisica Ennio De Giorgi, and I.N.F.N. - sezione di Lecce, Via Arnesano, I-73100 Lecce, Italy}

\affiliation[b]{Instituut voor Theoretische Fysica, KU Leuven.
	Celestijnenlaan 200D, B-3001 Leuven, Belgium \vspace{0.1cm}}
\affiliation[c]{Asia Pacific Center for Theoretical Physics, Postech, Pohang 37673, Korea }

\emailAdd{acbizet@gmail.com, marina.david@kuleuven.be,  alfredo.gonzalez@apctp.org}

\date{\today}
\abstract{Understanding how the thermodynamic properties of a black hole are modified when probed by D-branes is an important problem in AdS/CFT. This work focuses on a recently proposed black hole/D3-brane system in~AdS$_5\times$S$^5$, which is dual to four-dimensional~$\mathcal{N}=4$ SYM in the presence of a two-dimensional surface defect. The Laplace transform that extracts the asymptotic growth of states in this defect CFT naturally defines a thermodynamic approach in the gravitational side of the duality for which charges and entropy are real. Studying the superconformal defect index in a large-charge expansion for all values of~$N$, we compute the leading correction to the entropy of the combined system, which matches precisely with its gravity counterpart.
}

\begin{document} 
	\maketitle
	\flushbottom
 %%%%%%%%%%%%%%%%%%%%%%%%%%%%%%

\section{Introduction and Summary}
%%%%%%%%%%%%%%%%%%%%%%%%%%%%%%
%%%%%%%%%%%%%%%%%%%%%%%%%%%%%%

The AdS/CFT correspondence has allowed us to understand the Bekenstein-Hawking entropy for a large class of supersymmetric black holes as a semiclassical limit of the Boltzmann entropy of supersymmetric gauge theories \cite{Benini:2015eyy, Benini:2016rke,Choi:2018hmj, Cabo-Bizet:2018ehj, Benini:2018ywd, ArabiArdehali:2019tdm,Honda:2019cio,Cabo-Bizet:2019osg,Kim:2019yrz,Cabo-Bizet:2019eaf,Amariti:2019mgp,GonzalezLezcano2019, Lanir:2019abx,Goldstein:2019gpz,ArabiArdehali:2019orz, Murthy:2020rbd,PhysRevD.105.L021903,Agarwal:2020zwm, Benini:2020gjh,Cabo-Bizet:2020nkr, Cabo-Bizet:2021plf,Cassani:2021fyv,Jejjala:2021hlt,Jejjala:2022lrm,Aharony:2021zkr, Cabo-Bizet:2021jar,Goldstein:2020yvj, Choi:2021rxi,Choi:2023tiq}.  
Subleading corrections beyond the semiclassical result, which could be perturbative or logarithmic in the semi-classical expansion, and which could come from~$\alpha'$ corrections, as well as from other quantum effects, have been also computed and exactly matched across both sides of the duality, see for example~\cite{Bhattacharyya:2012ye,Liu:2017vbl,Liu:2017vll,David:2021eoq, GonzalezLezcano:2020yeb, Amariti:2020jyx,David:2021qaa,Cassani:2022lrk,Bobev:2021qxx, Bobev:2022bjm,Cassani:2023vsa, Bobev:2023dwx}. Despite these remarkable quantitatively precise advances, not much is understood yet about more drastic quantum gravity processes such as perturbing black holes with D-branes.

A pioneering attempt in this direction has been recently put forward in the context of AdS$_5$/CFT$_4$ in~\cite{Chen:2023lzq}. In this reference the authors studied the effect of perturbing a supersymmetric black hole in AdS$_5$~\cite{Chong:2005da,Chong:2005hr, Gutowski:2004ez, Gutowski:2004yv, Wu:2011gq,Kunduri:2006ek} with a~D$3$-brane\footnote{There are two non-trivial properties that need a closer analysis regarding how to preserve supersymmetry when inserting the D$3$-brane: The first is to check that it is possible for the D$3$-brane to be supersymmetric and the second is whether or not this supersymmetry is compatible with the one preserved by the black hole. Some indications that it is possible for this D$3$-brane to be supersymmetric have been presented in~\cite{Chen:2023lzq}. Here we assume this is true and leave a more rigorous check for future work. } extending across the time, radial, one compact direction in AdS$_5$ and one compact direction in the internal space~$S^5$. In the dual gauge theory, which is the four-dimensional~$SU(N)$ $\mathcal{N}=4$ SYM, inserting this probe D$3$-brane corresponds to inserting a surface operator \cite{Gukov:2006jk}, compatible with the supercharges used to construct the 4d superconformal index.

In the probe approximation, the authors of \cite{Chen:2023lzq} found that the free energy of the black hole/D$3$-brane system reduces to the sum of the free energy of the unperturbed black hole solution and the Dirac-Born-Infeld on-shell action of the D$3$-brane in the geometry of the unperturbed black hole solution, respectively. It may seem natural to assume that in the very same probe approximation the entropy of the total system reduces to the sum of the entropies of the two unperturbed components. Indeed, if one commits to this intuition then the charges of the D$3$-brane are fixed in terms of the charges of the embedding black hole solution. Unfortunately, the charges and entropy of the D$3$-brane fixed by this procedure turn out to be complex.\footnote{In contradistinction to the unperturbed system, in the presence of the D$3$-brane there is no non-linear constraint among real charges for which the extremal value of the corresponding entropy function becomes real and thus identifiable with the asymptotic value of a Boltzmann entropy.} As also stated in~\cite{Chen:2023lzq}, this result is intriguing, because the dual microstates that one counts in the field theory do indeed have real charges, and certainly their Boltzmann entropy is not complex. 
   
This naive contradiction strongly suggests that another procedure must be used to define the thermodynamic properties of the combined system. The goal of this paper is to find such a procedure. Indeed, we propose that the entropy of the system black hole/D$3$-brane is recovered by means of the natural holographic translation of the method used to count states in the holographic dual 4d-2d field theory system: the Laplace transform of the defect superconformal index.\footnote{This Laplace transforms only depends on the charges of the combined system and not on the charges of its individual components.}

For the case of the superconformal index without the insertion of the surface defect, the Laplace transform -- in the leading order in the semiclassical large charge approximation -- picks up two leading complex conjugated saddle points whose contributions add up to give a real entropy  \cite{Agarwal:2020zwm, Cabo-Bizet:2020ewf, Beccaria:2023hip}. Similarly, as we show here, we find this also to be the case when the defect is introduced in the system. In the gravitational picture, these leading saddle points correspond to two complex geometries that serve as saddle points of the Euclidean gravitational path integral. Borrowing the field-theory procedure to the holographic dual setup implicitly defines how to compute the corrections to the entropy that the D$3$-brane induces when probing the black hole.
   
Using a Cardy-like expansion we confirm the results of~\cite{Chen:2023lzq} for the free energy of the 4d-2d field theory, and extend them, both to finite~$N$ and beyond the probe approximation. We find that at leading order in the Cardy-like expansion, inserting the defect does not change the shape of the saddle point governing the growth of the unperturbed 4d superconformal index. Surprisingly, this tells us that the naive probe approximation is sufficiently precise to exactly describe the 4d-2d system at large charges. In the string-theory side of the duality, this result predicts a fully backreacted answer for the entropy of the perturbed black hole at leading order in the Cardy-like expansion. It would be very interesting to understand whether the~$\frac{1}{N}$ corrections induced by the presence of the D$3$-brane can be understood, geometrically, as a change in the area of the horizon. In order to answer this question we would need to understand how the D$3$-brane backreacts the geometry in the bulk. We leave that problem for future work.

The paper is organized as follows. Section~\ref{section:gravity} summarizes the supergravity theory, and other useful background information. We motivate and present our prescription to study the thermodynamics of the combined black hole/D$3$-brane system and perform the Laplace transform to extract the microcanonical entropy. In Section~\ref{section:4dindex} we study the field theory dual description and we revisit the 4d computation in the absence of defect and recall how to evaluate the asymptotic growth of the index in that case. In Section~\ref{sec:defectIndex} we compute the defect index. We analyze the 2d index of the surface defect at large charges by implementing a systematic Cardy-like expansion. In Section~\ref{section:conclusions} we conclude with brief remarks and questions for the future. Appendix~\ref{appendix:reviewofpreviousapproach} reviews the thermodynamic procedure used in~\cite{Chen:2023lzq} while Appendix~\ref{App:elliptic:functions} summarizes useful mathematical identities.

%%%%%%%%%%%%%%%%%%%%%%%%%%%%%%
%%%%%%%%%%%%%%%%%%%%%%%%%%%%%%
\section{The gravity theory} \label{section:gravity}
%%%%%%%%%%%%%%%%%%%%%%%%%%%%%%
%%%%%%%%%%%%%%%%%%%%%%%%%%%%%%

In this section, we summarize the five-dimensional gauged black hole solution of \cite{Chong:2005hr} and its thermodynamic properties. After reviewing its supersymmetric limit we carry on to study the effects of adding the probe D$3$-brane. Then we proceed to study the thermodynamics of the combined system.
%%%%%%%%%%%%%%%%%%%%%%%%%%%%%%%%%%%%%%%
%%%%%%%%%%%%%%%%%%%%%%%%%%%%%%%%%%%%%%%
\subsection{The black hole solution} \label{sec:ugraRev}
%%%%%%%%%%%%%%%%%%%%%%%%%%%%%%%%%%%%%%%
%%%%%%%%%%%%%%%%%%%%%%%%%%%%%%%%%%%%%%%
We consider five-dimensional minimal gauged supergravity whose action takes the form
\begin{align} \label{eq:5daction}
	S &= \frac{1}{16\pi G_{5}}\int d^5 x \left[(R + 12g^2)\star 1 - \frac{2}{3g^2}F\wedge \star F + \frac{8}{27g^3}F \wedge F \wedge A \right],
\end{align}
where $F=dA$ and $g$ is the inverse length of AdS. The five-dimensional coordinates describing the solution are $\{t,r,\theta,\phi,\psi\}$ where $0\leq\phi,\psi \leq 2\pi$ and  $ 0 \leq \theta \leq \frac{\pi}{2}$. The equations of motion can be derived from \eqref{eq:5daction}
\begin{align}
    0 &= R_{\mu\nu}-\frac{1}{2}g_{\mu\nu}R - 6 g^2 g_{\mu\nu} - \frac{4}{3g^2}\left(\frac{1}{2}F^2_{\mu\nu}-\frac{1}{8}g_{\mu\nu}F^2\right), \qquad 0 = d\star F + \frac{2}{3g}F\wedge F.
\end{align}
We review the known nonextremal nonsupersymmetric black hole solution with one electric charge and two rotations as was studied in \cite{Chong:2005hr}. The solution of the metric and gauge field are given by
\begin{equation}
	\begin{aligned}
		d s^2_{\text{AdS}_5}= & -\frac{\Delta_\theta\left[\left(1+g^2 r^2\right) \rho^2 d t+2 q \nu\right] d t}{\Xi_a \Xi_b \rho^2}+\frac{2 q \nu \omega}{\rho^2}+\frac{f}{\rho^4}\left(\frac{\Delta_\theta d t}{\Xi_a \Xi_b}-\omega\right)^2+\frac{\rho^2 d r^2}{\Delta_r}+\frac{\rho^2 d \theta^2}{\Delta_\theta} \\
		& +\frac{r^2+a^2}{\Xi_a} \sin ^2 \theta d \phi^2+\frac{r^2+b^2}{\Xi_b} \cos ^2 \theta d \psi^2, \\
		A= & \frac{3 q}{2\rho^2}\left(\frac{\Delta_\theta d t}{\Xi_a \Xi_b}-\omega\right)+\alpha_5 d t,
	\end{aligned}
\end{equation}	
where we have added a pure gauge term $\alpha_{5}dt$ with $\alpha_5$ being a constant. The remaining functions in the metric and 1-form are
\begin{equation}
	\begin{aligned}
		\nu & =b \sin ^2 \theta d \phi+a \cos ^2 \theta d \psi, \quad &\Delta_r & =g^2\left(r^2+a^2\right)\left(r^2+b^2\right)\left(1+\frac{1}{g^2r^2}\right)+ \frac{q^2+2 a b q}{r^2}-2 m, \\
		\omega & =a \sin ^2 \theta \frac{d \phi}{\Xi_a}+b \cos ^2 \theta \frac{d \psi}{\Xi_b}, \quad &\Delta_\theta & = 1-a^2 g^2 \cos ^2 \theta-b^2 g^2 \sin ^2 \theta,
		\\
		\rho^2 & = r^2+a^2 \cos ^2 \theta+b^2 \sin ^2 \theta, \quad & \Xi_a & = 1-a^2 g^2,\\
		f & = 2 m \rho^2-q^2+2 a b q g^2 \rho^2,
        \quad & \Xi_b & = 1-b^2 g^2.
	\end{aligned}
\end{equation}
For the general non-extremal solution with no supersymmetry, there are four independent parameters that characterize the black hole
\begin{align}
	\{a,b,m,q\}.
\end{align}
Moreover, we may sometimes find it convenient to swap one of the parameters, namely, $q$ with the horizon radius, i.e.,
\begin{align} \label{eq:qrelation}
    q = - a b \pm r_+ \sqrt{-a^2 \left(b^2 g^2+g^2 r_+^2+1\right)-b^2 \left(g^2 r_+^2+1\right)-g^2 r_+^4+2 m-r_+^2}.
\end{align}
The electric charges and angular momentum can be computed via the Komar integrals
\begin{align} \label{eq:CCLP-Q}
	Q_{\text{BH}} &= \frac{1}{16\pi G_{5}}\int_{S^3}\left(\frac{4}{3g^2}\right)\star F - \frac{8}{9g^3} F \wedge A = \frac{1}{G_{5}}\frac{\pi q}{2 g \Xi_a \Xi_b},
    \\ \label{eq:CCLP-Ja}
    J_{1,\text{BH}} &= \frac{1}{16\pi G}\int_{S^3} \star d \xi_{\phi} = \frac{1}{G_{5}}\frac{\pi\left[2 a m+q b\left(1+a^2 g^2\right)\right]}{4 \Xi_a^2 \Xi_b},
    \\ \label{eq:CCLP-Jb}
    J_{2,\text{BH}} &= \frac{1}{16\pi G}\int_{S^3} \star d \xi_{\psi} = \frac{1}{G_{5}}\frac{\pi\left[2 b m+q a\left(1+b^2 g^2\right)\right]}{4 \Xi_b^2 \Xi_a},
\end{align}
where $\xi_{\phi}$ and $\xi_{\psi}$ are dual to Killing vector $-\partial_{\phi}$ and $-\partial_{\psi}$ respectively such that
\begin{align}
    %\xi^{\mu}_{1}\partial_{\mu} = -\partial_{\phi}, \xi^{\mu}_{2}\partial_{\mu} = -\partial_{\psi}
    %\\
    \xi_{\phi} = - g_{\mu\phi} dx^{\mu}, \qquad \xi_{\psi} = - g_{\mu\psi} dx^{\mu}.
\end{align}
The charges are evaluated at the asymptotic boundary and for this reason, the Chern Simons term in the integral for the electric charge does not contribute to the charge. In fact there are different notions of charge and we refer the reader to \cite{Marolf:2000cb} for more details. The energy can be found from the AMD formalism
\begin{equation} \label{eq:CCLP-mass}
	E_{\text{BH}} = \frac{1}{G_{5}}\frac{m \pi\left(2 \Xi_a+2 \Xi_b-\Xi_a \Xi_b\right)+2 \pi q a b g^2\left(\Xi_a+\Xi_b\right)}{4 \Xi_a^2 \Xi_b^2} .
\end{equation}
The Hawking temperature is derived by requiring appropriate periodic identifications in Euclidean time, which leads us to
\begin{align} \label{eq:Tgeneral}
    T_{\text{BH}} = \beta^{-1}_{\text{BH}} &= \frac{r_{+}^4\left[\left(1+g^2\left(2 r_{+}^2+a^2+b^2\right)\right]-(a b+q)^2\right.}{2 \pi r_{+}\left[\left(r_{+}^2+a^2\right)\left(r_{+}^2+b^2\right)+a b q\right]}.
\end{align}

The angular velocities $\Omega_{1}$ and $\Omega_{2}$ are found to be
%which we find to be 
\begin{align} \label{eq:Omegageneral}
    \Omega_{1,\text{BH}} & =\frac{a\left(r_{+}^2+b^2\right)\left(1+g^2 r_{+}^2\right)+b q}{\left(r_{+}^2+a^2\right)\left(r_{+}^2+b^2\right)+a b q}, \quad
    \Omega_{2,\text{BH}} = \frac{b\left(r_{+}^2+a^2\right)\left(1+g^2 r_{+}^2\right)+a q}{\left(r_{+}^2+a^2\right)\left(r_{+}^2+b^2\right)+a b q}.
\end{align}
We can now define the null Killing vector field
\begin{align}
    \chi^{\mu} = \partial_{t} + \Omega_1 \partial_{\phi} + \Omega_2 \partial_{\psi},
\end{align}
and the electrostatic potential is
\begin{align} \label{eq:Phigeneral}
	\Phi_{\text{BH}} &= \left. \chi^{\mu} A_{\mu} \right|_{r\to r_+} - \left.\chi^{\mu} A_{\mu} \right|_{r\to r_+} = \frac{3 g q r_{+}^2}{2\left(\left(r_{+}^2+a^2\right)\left(r_{+}^2+b^2\right)+a b q\right)}.
\end{align}
The entropy can be computed via the area of the horizon
\begin{equation} \label{eq:CCLP-S}
	S_{\text{BH}} = \frac{1}{G_{5}}\frac{\pi^2\left[\left(r_{+}^2+a^2\right)\left(r_{+}^2+b^2\right)+a b q\right]}{2 \Xi_a \Xi_b r_{+}}.
\end{equation}
Once we have computed these thermodynamic quantities, we may deduce the on-shell action from the quantum statistical relation
\begin{align}
\begin{split}
    I_{\text{BH}} &= \beta_{\text{BH}} E_{\text{BH}} - S_{\text{BH}} - \beta_{\text{BH}} \Omega_{1,\text{BH}} J_{1,\text{BH}} - \beta_{\text{BH}} \Omega_{2,\text{BH}} J_{2,\text{BH}} - \beta_{\text{BH}} \Phi_{\text{BH}} Q_{\text{BH}}
    \\&=
    \frac{\pi  \beta}{4 G_{5} \Xi_a \Xi_b}  \left(m-g^2 \left(a^2+r_{+}^2\right) \left(b^2+r_{+}^2\right)-\frac{q^2 r_{+}^2}{\left(a^2+r_{+}^2\right) \left(b^2+r_{+}^2\right)+a b q}\right).
\end{split}
\end{align}
%%%%%%%%%%%%%%%%%%%%%%%%%%%%%%%%%%%%%%%%%%%%%%%%%%%%%%%%%%%%%%%%%%%%%%%%%%%%
%%%%%%%%%%%%%%%%%%%%%%%%%%%%%%%%%%%%%%%%%%%%%%%%%%%%%%%%%%%%%%%%%%%%%%%%%%%%
\subsection{The supersymmetric limit} \label{sec:susylimit}
%%%%%%%%%%%%%%%%%%%%%%%%%%%%%%%%%%%%%%%%%%%%%%%%%%%%%%%%%%%%%%%%%%%%%%%%%%%%
%%%%%%%%%%%%%%%%%%%%%%%%%%%%%%%%%%%%%%%%%%%%%%%%%%%%%%%%%%%%%%%%%%%%%%%%%%%%

We are interested in solutions that admit a Killing spinor, i.e., preserve $\mathcal{N}=2$ supersymmetry. The BPS bound
\begin{align} \label{eq:CCLP-SUSYbound1}
	E_{\text{BH}} = g J_{1,\text{BH}} + g J_{2,\text{BH}} + \frac{3}{2}g Q_{\text{BH}}\,,
\end{align}
is saturated for
\begin{equation} \label{eq:CCLP-constraint1}
	q=\frac{m}{1+(a+b) g}.
\end{equation}
This can be found by imposing \eqref{eq:CCLP-Q}, \eqref{eq:CCLP-Ja}, \eqref{eq:CCLP-Jb} and \eqref{eq:CCLP-mass} into \eqref{eq:CCLP-SUSYbound1}. Keeping in mind \eqref{eq:qrelation}, we find that the parameter $q$ simplifies to the following 
\begin{align} \label{eq:qSUSY}
    \begin{split}
        q &= -a b+a g r_+^2+b g r_+^2+r_+^2  \pm i r_+ \left(a b g+a+b-g r_+^2\right)
        \\&= (a - n_0 i r_{+})(b - n_0 i r_{+})(-1 + n_0 igr_+).
    \end{split}
\end{align}
From now on, we denote $n_0=\pm 1$ for the upper/lower sign in \eqref{eq:qSUSY} respectively. This choice of sign can be interpreted as a choice in one of two branches that dominate the path integral and denote a growth of states. We shall come back to this point in great detail in Section~\ref{subsec:ExtremizationGravity}. In the supersymmetric limit, the temperature, angular velocities and electrostatic potential in \eqref{eq:Tgeneral}, \eqref{eq:Omegageneral} and \eqref{eq:Phigeneral} are complex
\begin{align}
    T_{\text{BH}} &= \frac{g \left(r_+^2-r_*^2\right) \left(2 r_+ (a g+b g+1)+i n_0 g \left(r_*^2-3 r_+^2\right)\right)}{2 \pi  \left(a-i n_0 r_+\right) \left(b-i n_0 r_+\right) \left(g r_*^2+i n_0 r_+\right)},
    \\
    \Omega_{1,\text{BH}} &= \frac{g \left(a r_+ - i n_0  r_*^2\right) \left(1 - i n_0 g r_+\right)}{\left(a -  i n_0 r_+\right) \left(r_+ - i n_0 g r_*^2\right)},
    \\
    \Omega_{2,\text{BH}} &=  \frac{g \left(b r_+ - i n_0  r_*^2\right) \left(1 - i n_0 g r_+\right)}{\left(b -  i n_0 r_+\right) \left(r_+ - i n_0 g r_*^2\right)},
    \\
    \Phi_{\text{BH}} &= \frac{3 g r_+ \left(1-i n_0 g r_+\right)}{2 r_+-2 i n_0 g r_*^2},
\end{align}
and we can equivalently find a linear constraint among the angular velocities and electrostatic potentials of the black hole
\begin{align} \label{eq:relationbetan0}
    \beta_{\text{BH}} \left(g+\Omega_{1,\text{BH}}+\Omega_{2,\text{BH}}-2\Phi_{\text{BH}} \right) = 2\pi i n_0.
\end{align}
Once \eqref{eq:CCLP-SUSYbound1} is imposed, we find that if we want to preserve the reality of the parameters $a,b,q$ and $m$, we find that
\begin{align} \label{eq:CCLP-constraint2}
	r^{\star} = \sqrt{\frac{a+b+abg}{g}},
\end{align}
and this is the exact value where the discriminant of $r^2 \Delta_r$ is zero, i.e., the inner and outer horizons coincide and we land in the extremal regime of the solution. This analysis leads us to conclude that supersymmetric Lorentzian solutions must be extremal if we preserve the reality of roots of $\Delta_r$ to avoid naked singularities.

Imposing both these conditions \eqref{eq:CCLP-constraint1} and \eqref{eq:CCLP-constraint2} leads to the following thermodynamic relations
\begin{equation}
	\begin{aligned}
		Q^{\star}_{\text{BH}} = & -\frac{1}{G_{5}}\frac{\pi (a+b)}{2 g(1-a g)(1-b g)},
		\\
		J^{\star}_{1, \text{BH}} = & \frac{1}{G_{5}}\frac{\pi(a+b)(2 a+b+a b g)}{4 g(1-a g)^2(1-b g)}, \\
		J^{\star}_{2, \text{BH}} = & \frac{1}{G_{5}}\frac{\pi(a+b)(a+2 b+a b g)}{4 g(1-a g)(1-b g)^2},
		\\
		E^{\star}_{\text{BH}} = & \frac{1}{G_{5}}\frac{\pi(a+b)}{4 g(1-a g)^2(1-b g)^2}((1-a g)(1-b g) + (1+a g)(1+b g)(2-a g-b g)), \\
		S^{\star}_{\text{BH}} = & \frac{1}{G_{5}}\frac{\pi^2(a+b) \sqrt{a+b+a b g}}{2 g^{3 / 2}(1-a g)(1-b g)}, 
	\end{aligned}
\end{equation}
which are now all real-valued expressions.
We shall call the BPS limit the limit of the solution where both extremal and supersymmetric conditions are imposed and we denote this by $\star$. The family of solutions has now been reduced to two free parameters, $a$ and $b$. Revisiting the quantum statistical relation, we introduce the variables
\begin{align} \label{eq:omegavarphivariables}
    \omega_{1,\text{BH}} &= \frac{\beta_{\text{BH}}}{2\pi i} (\Omega_{1,\text{BH}} - \Omega^{\star}_{1,\text{BH}}), 
    \quad \omega_{2,\text{BH}} = \frac{\beta_{\text{BH}}}{2\pi i} (\Omega_{2,\text{BH}} - \Omega^{\star}_{2,\text{BH}}), 
    \quad \frac{3}{2}\varphi_{\text{BH}} = \frac{\beta_{\text{BH}}}{2\pi i} (\Phi_{\text{BH}} - \Phi^{\star}_{\text{BH}}),
\end{align}
with
\begin{align}
    \Omega^{\star}_{1,\text{BH}} = g, \qquad \Omega^{\star}_{2,\text{BH}} = g, \qquad \Phi^{\star}_{\text{BH}} = \frac{3g}{2}.
\end{align}
Note that the BPS values of the chemical potentials are independent of which saddle we consider. Imposing these new variables into \eqref{eq:relationbetan0}, we find the new linear constraint among the chemical potentials takes the form
\begin{align} \label{eq:linearconstraint}
    \omega_{1,\text{BH}} + \omega_{2,\text{BH}} - 3\varphi_{\text{BH}} = n_0.
\end{align}
With some manipulation, as the supersymmetric limit must be taken carefully, the quantum statistical relation can now be rewritten as a statement independent of the temperature
\begin{align}
    \begin{split}
        I_{\text{BH, SUSY}} &= - S_{\text{BH}} - \omega_{1,\text{BH}} J_{1,\text{BH}} - \omega_{2,\text{BH}}  J_{2,\text{BH}} - \varphi_{\text{BH}} Q_{\text{BH}} 
        = \frac{\pi^2 i}{2g^3 G_5}\frac{\varphi^3_{\text{BH}}}{\omega_{1,\text{BH}} \omega_{2,\text{BH}}},
    \end{split}
\end{align}
and via the holorgraphic dictionary, we arrive at the following on-shell action
\begin{align}
     I_{\text{BH, SUSY}} = \pi i N^2\frac{\varphi^3_{\text{BH}}}{\omega_{1,\text{BH}} \omega_{2,\text{BH}}}.
\end{align}

%%%%%%%%%%%%%%%%%%%%%%%%%%%%%%%%%%%%%%%%%%%%%%%%%%%%%%%%%%%%%%
%%%%%%%%%%%%%%%%%%%%%%%%%%%%%%%%%%%%%%%%%%%%%%%%%%%%%%%%%%%%%%
\subsection{The combined system: black hole and probe D$3$-brane} \label{sec:thermodynBHD3}
%%%%%%%%%%%%%%%%%%%%%%%%%%%%%%%%%%%%%%%%%%%%%%%%%%%%%%%%%%%%%%
%%%%%%%%%%%%%%%%%%%%%%%%%%%%%%%%%%%%%%%%%%%%%%%%%%%%%%%%%%%%%%

Next we move on to introduce the methodology we follow to compute the $\mathcal{O}(N)$ corrections to the entropy induced by the probe D$3$-brane. It is important to emphasize that even when~$\alpha'$ corrections are included, the entropy of the supersymmetric black hole receives corrections of~$\mathcal{O}(N^0)$ \cite{Bobev:2021qxx,Bobev:2022bjm,Cassani:2022lrk,Cassani:2023vsa} and so for the purposes of this paper, they can be ignored.

Assuming unequal black hole angular momentum and equal black hole electric charges,~\cite{Chen:2023lzq} found that the supersymmetric on-shell action of the D$3$-brane is
\begin{align}\label{eq:D3braneOnshellAction}
    I_{\text{D3, SUSY}} &= - 2\pi i N \frac{\widetilde{\varphi}^2}{\widetilde{\omega}_1}.
\end{align}
This result comes from regularizing the Dirac-Born-Infeld and the Wess-Zumino contributions~\cite{Chen:2023lzq}.

The wide tilde denotes the variables in the perturbed system and not just of the unperturbed black hole, e.g., $\widetilde{\varphi}\neq {\varphi}_{\text{BH}}\,$.

To understand physically the dependence on the chemical potentials in the on-shell action, let us review how the brane is extended into the bulk. On the AdS$_5$ coordinates, the coordinates~$\theta$ and~$\psi$ are fixed
\begin{align}
    \text{AdS}_5:& \quad (t,r,\theta=\theta_0,\phi,\psi=\psi_0),
\end{align}
while on the $S^5$, the D$3$-brane only wraps around one of the coordinates while the others remain fixed
\begin{align}
    \text{S}^5:& \quad (\phi_1,\phi_2=\phi_{2,0},\phi_3 = \phi_{3,0},\bar\theta=\bar\theta_0, \bar \psi=\bar\psi_0).
\end{align}
As the angular momentum of the system comes from the symmetries associated to the Killing vectors $\partial_{\phi}$ and $\partial_{\psi}$, only one is set to be free which means that the on-shell action may only depend on the chemical potential conjugate to the angular momentum associated to $\partial_{\phi}$. On the other hand, the electric charges come from the~$S^5$, in particular, from~$\phi_i$ and we expect that the on-shell action depends on the two potentials associated to the electric charges from the fixed angles $\phi_2$ and $\phi_3$.

Although we have studied the black hole at equal electric charges and potentials, once the D$3$-brane is introduced into the system, the potentials acquire a subleading correction in the~$1/N$-expansion. Denoting the perturbed potentils as~$\widetilde\varphi_1, \widetilde\varphi_2$ and $\widetilde\varphi_3$ we expect the on-shell actions for the fully refined system to be
\be
\widetilde I\,=\,I_{\text{BH, SUSY}}\,+\,I_{\text{D3, SUSY}},
\ee
where
\begin{align} \label{eq:Igeneral}
    I_{\text{BH, SUSY}} &= \pi i N^2\frac{\widetilde \varphi_1 \widetilde \varphi_2 \widetilde \varphi_3}{\widetilde\omega_1 \widetilde\omega_2}, \qquad I_{\text{D3, SUSY}} = - 2\pi i N \frac{\widetilde \varphi_2 \widetilde \varphi_3}{\widetilde \omega_1}.
\end{align}
This expectation is reassured by microscopic computations of the fully refined 4d superconformal index \cite{Benini:2018ywd, Choi:2018hmj, Cabo-Bizet:2018ehj} and our calculation of the 2d index in Section~\ref{section:2dindex}. To prove \eqref{eq:Igeneral}, one would need to study the BPS limit of the fully refined AdS$_5$ black hole solution of \cite{Wu:2011gq}.

The quantum statistical relation of the perturbed system is
\begin{align}
    \widetilde I &= - \widetilde S - \textstyle{\sum_{k=1}^2}\widetilde \omega_{k} \widetilde J_{k} - \textstyle{\sum_{I=1}^3}\widetilde\varphi_{I} \widetilde Q_{I}.
\end{align}

%%%%%%%%%%%%%%%%%%%%%%%%%%%%%%%%%%%%%
%%%%%%%%%%%%%%%%%%%%%%%%%%%%%%%%%%%%%
\subsection{The extremization} \label{subsec:ExtremizationGravity}
%%%%%%%%%%%%%%%%%%%%%%%%%%%%%%%%%%%%%
%%%%%%%%%%%%%%%%%%%%%%%%%%%%%%%%%%%%%
The entropy of the system can be found by extremizing the entropy function
\begin{align} \label{eq:extremeq}
	\widetilde{S} = -\widetilde{I} - 2\pi i \sum_{I=1}^{3} \widetilde{\varphi}_{I} \widetilde{Q}_{I} - 2\pi i \sum_{k=1}^{2} \widetilde{\omega}_{k}\widetilde{J}_{k} + 2\pi i  \widetilde{\Lambda} \left(\sum_{I=1}^{3} \widetilde{\varphi}_I - \sum_{k=1}^2 \widetilde{\omega}_k + n_0\right).
\end{align}
We impose the linear constraint among the chemical potentials via the Lagrange muliplier $\Lambda$.\footnote{We note that in the gauge theory side the two choices~$n_0=\pm1$ correspond to two different saddle points of a multi-dimensional Laplace transform used to exchange from canonical to microcanonical ensemble as well as to impose the gauge-singlet constraint. We will further elaborate on this in Section~\ref{sec:defectIndex}.}

The extremization of~\eqref{eq:extremeq} leads to the following constraints
\begin{align} \label{eq:extremizationDelta}
		0 &= 2\pi i ( \widetilde{\Lambda} - \widetilde{Q}_{I}) - i \pi N^2 \frac{\widetilde{\varphi}_1 \widetilde{\varphi}_2 \widetilde{\varphi}_3}{\widetilde{\omega}_1 \widetilde{\omega}_2 \widetilde{\varphi}_I} + 2i \pi N \frac{\widetilde{\varphi}_2 \widetilde{\varphi}_3}{\widetilde{\omega}_1 (\varphi_I^2 \widetilde{\varphi}_2 + \varphi_I^3 \widetilde{\varphi}_3)},  &\qquad I&=1,2,3,
		\\ \label{eq:extremizationomega}
		0 &= - 2\pi i ( \widetilde{\Lambda} + \widetilde{J}_{k}) + i \pi N^2 \frac{\widetilde{\varphi}_1 \widetilde{\varphi}_2 \widetilde{\varphi}_3}{\widetilde{\omega}_1 \widetilde{\omega}_2 \widetilde{\omega}_k} - 2i \pi N \frac{\widetilde{\varphi}_2 \widetilde{\varphi}_3}{\delta_k^1(\widetilde{\omega}_1)^2}, &\qquad k&=1,2\,.
\end{align}
Solving for the charges and substituting back into~\eqref{eq:extremeq} we find the extremal value of the entropy function
\begin{align} \label{eq:entropyLambda}
	\widetilde{S} = 2\pi i n_0  \widetilde{\Lambda}\,,
\end{align}
which has the same structure as the result obtained in the absence of the probe D$3$-brane, although the value of~$\widetilde{\Lambda}$ as a function of charges changes for the perturbed system.

In order to identify~\eqref{eq:entropyLambda} with the entropy, which is a real-valued quantity, one may constrain the charges of the system to the locus~$\text{Im}(\widetilde{S})=0$. In the absence of the D$3$-brane, this is the well-known non-linear constraint among charges that happens to be equivalent to the vanishing of the Bekenstein-Hawking temperature. In the presence of the D$3$-brane and for real charges, there is no solution to the locus~$\text{Im}(\widetilde{S})=0\,$. As explained in the introduction, the field-theoretic analysis will provide the solution to this puzzle: at large charges of order~$N^2$ for large~$N$ the entropy of the system approaches, asymptotically, to the real part of entropy functional~$\widetilde{S}$ of the dominating saddle points, 
\be
\text{Entropy}\,\sim\, \text{Re}(\widetilde{S})\,,
\ee
without the need of imposing further constraints on the charge locus. This result comes from the addition of contributions coming from two complex conjugated saddles, each of them with its own on-shell entropy functional. This addition yields a real-valued asymptotic entropy. Let us elaborate. With four of the extremization equations in \eqref{eq:extremizationDelta} and \eqref{eq:extremizationomega}, we solve for $\widetilde{\varphi}_{1},\widetilde{\varphi}_{2},\widetilde{\varphi}_{3}$, and $\widetilde{\omega}_2$
\begin{align}
    \begin{split}
        \widetilde\varphi_1 &= \frac{2\widetilde\omega_1 (\widetilde{J}_{1} N+  \widetilde{\Lambda}  ( \widetilde{\Lambda} + N- \widetilde{Q}_{3})+\widetilde{Q}_{2} (\widetilde{Q}_{3}- \widetilde{\Lambda} ))^2}{N^2 (\widetilde{J}_{2}+ \widetilde{\Lambda} ) (\widetilde{Q}_{2}- \widetilde{\Lambda} ) (\widetilde{Q}_{3}- \widetilde{\Lambda} )},
        \\
        \widetilde\varphi_2 &= -\frac{\widetilde\omega_1 (\widetilde{J}_{1}+ \widetilde{\Lambda} )}{\widetilde{Q}_{2}- \widetilde{\Lambda} },
        \\
        \widetilde\varphi_3 &= -\frac{\widetilde\omega_1 (\widetilde{J}_{1}+ \widetilde{\Lambda} )}{\widetilde{Q}_{3}- \widetilde{\Lambda} },
        \\
        \widetilde\omega_2 &= \frac{\widetilde\omega_1 (\widetilde{J}_{1}+ \widetilde{\Lambda} ) (\widetilde{J}_{1} N+ \widetilde{\Lambda}  ( \widetilde{\Lambda} + N- \widetilde{Q}_{3})+\widetilde{Q}_{2} (\widetilde{Q}_{3}- \widetilde{\Lambda} ))}{(\widetilde{J}_{2}+ \widetilde{\Lambda} ) ( \widetilde{\Lambda} -\widetilde{Q}_{2}) ( \widetilde{\Lambda} -\widetilde{Q}_{3})}.
    \end{split}
\end{align}
Note that the dependence on~$\widetilde \omega_1$ is trivial. Imposing these relations into the remaining extremization equation in \eqref{eq:extremizationDelta} and \eqref{eq:extremizationomega} leads to the following equation for~$\widetilde{\Lambda}$
\begin{align} \label{eq:cubiceq}
    0 &= 2( \widetilde{\Lambda} -\widetilde{Q}_1)( \widetilde{\Lambda} -\widetilde{Q}_2)( \widetilde{\Lambda} -\widetilde{Q}_3) - N^2 ( \widetilde{\Lambda} + \widetilde{J}_1)( \widetilde{\Lambda} +\widetilde{J}_2) + 2N ( \widetilde{\Lambda} + \widetilde{J}_1)( \widetilde{\Lambda} -\widetilde{Q}_1).
\end{align}
From \eqref{eq:cubiceq}, we can see that the first two terms have the same form as the cubic equation for $\Lambda$ for the black hole in absence of the D$3$-brane, but now the charges correspond to the combined system. The last term in \eqref{eq:cubiceq} is of order $\mathcal{O}(N)$ and can be treated perturbatively. To keep track of the leading terms in $N$, we consider a rescaling of the form

\begin{align} \label{eq:LambdaJQexpansion}
	 \widetilde{\Lambda} &= N^2 \Lambda, \quad
	\widetilde{J}_{k} = N^2 J_{k,\text{BH+D3}}, \quad
	\widetilde{Q}_{I} = N^2 Q_{I,\text{BH+D3}}.
\end{align}
To ease presentation we remove the subscript ``BH+D3" in the remaining of this section. Then, we find

\begin{align} \label{eq:cubiceq1}
    \begin{split}
        0 &= 2(\Lambda -Q_{1 })(\Lambda -Q_{2 })(\Lambda -Q_{3 }) - (\Lambda + J_{1 })(\Lambda +J_{2 })
        + \frac{2}{N} (\Lambda + J_{1 })(\Lambda -Q_{1 }).
    \end{split}
\end{align}
Moreover, the first two terms, with the probe D$3$-brane turned off, is the usual cubic equation for $\Lambda$ that appears when only considering the black hole \cite{Cabo-Bizet:2018ehj,Choi:2018hmj,Benini:2018ywd}. We analyze the cubic equation in a perturbative expansion in $N$ by first considering the general form of the cubic polynomial
\begin{align} \label{eq:cubicBH}
   P(a_\ell,\Lambda) \equiv a_0 + a_1 \Lambda + a_2 \Lambda^2 + a_3 \Lambda^3 = (\Lambda-\Lambda_{+})(\Lambda-\Lambda_{0})(\Lambda-\Lambda_{-}) & =0\,, \quad a_{0,\,1, \, 2,\,3} \, \in \mathbb{R}\, , 
 \end{align}
where $\Lambda_{\pm}$ and $\Lambda_{0}$ are the three roots of the cubic equation.

With the combined system, each coefficient in the polynomial as well as the Lagrange multiplier may also receive corrections
\begin{align}
    a_\ell = a_\ell^{(0)} + \frac{1}{N}a_{\ell}^{(1)}, \qquad \Lambda = \Lambda^{(0)} + \frac{1}{N} \delta \Lambda^{(1)}, \qquad \ell=0,1,2,3.
\end{align}
The given values of $a_\ell^{(0)}$ and $a_\ell^{(1)}$ are
\begin{subequations}
 \begin{align} \label{eq:abcdBH}
         a^{(0)}_0 & =-2 Q_{1 }  Q_{2 }  Q_{3 }-  J_{1 } J_{2 }, & a^{(1)}_{0} &= 2 J_{1 }Q_{1 }, \quad   \\
         a^{(0)}_1 & = 2 ( Q_{1 }  Q_{2 }, +  Q_{2 }  Q_{3 } +  Q_{1 }  Q_{3 }) -  \textstyle\sum_{k=1}^2 J_{k},  & a^{(1)}_{1} &= 2 (Q_{1 } - J_{1 }), \quad \\
         a^{(0)}_2 & = -2 \textstyle\sum_{I=1}^3 Q_{I } - 1, & a^{(1)}_{2} &= -2, \\
         a^{(0)}_3 & = 2, & a^{(1)}_3 &= 0.\,
 \end{align}
\end{subequations}

We can expect that the roots of \eqref{eq:cubicBH} are shifted by a subleading correction in $N$
\begin{align} \label{eq:perturbedcubiceq}
    \sum_{l=0}^3 \left(a^{(0)}_l+ \frac{1}{N} a^{(1)}_l\right)\left(\Lambda^{(0)} + \frac{1}{N}\Lambda^{(1)}\right)^l & = 0.
\end{align}
Expanding for large $N$, we find
\begin{align}
    P(a_{\ell},\Lambda) = P(a^{(0)}_{\ell}, \Lambda^{(0)}) + \frac{1}{N}\sum_{\ell=0}^3 \frac{\partial P(a^{(0)}_{\ell}, \Lambda^{(0)})}{\partial a_{\ell}}a^{(1)}_{\ell} + \frac{1}{N}\frac{\partial P(a^{(0)}_{\ell}, \Lambda^{(0)})}{\partial \Lambda} \Lambda^{(1)} + \mathcal{O}(N^{-2}).
\end{align}
Note that the expansion is only valid up to $\mathcal{O}(N^{-1})$ as higher corrections must also be supplemented, for example, by higher derivative corrections from the black hole. Evaluating at the roots $\Lambda^{(0)}_{k=\pm, 0}$, the zeroth order term vanishes, as expected, while the subleading terms are in general nonzero
\begin{subequations}
\begin{align} \label{eq:Pderivativea}
    \sum_{\ell=0}^3 \frac{\partial P(a^{(0)}_{\ell}, \Lambda^{(0)}_{k})}{\partial a_{\ell}}a^{(1)}_{\ell} &= a^{(1)}_{0} + a^{(1)}_{1}\Lambda^{(0)}_{k}+a^{(1)}_{2}(\Lambda^{(0)}_{k})^2+a^{(1)}_{3}(\Lambda^{(0)}_{k})^3,
    \\
    \begin{split} \label{eq:PderivativeLambda}
    \frac{\partial P(a^{(0)}_{\ell}, \Lambda^{(0)}_k)}{\partial \Lambda}&= a^{(0)}_{1}+2a^{(0)}_{2}\Lambda^{(0)}_{k}+3a^{(0)}_{3}(\Lambda^{(0)}_{k})^2 \\&= \frac{1}{2}\sum_{m \neq n \neq k}(\Lambda^{(0)}_{k}-\Lambda^{(0)}_{m})(\Lambda^{(0)}_{k}-\Lambda^{(0)}_{n}).%(1-\delta^{k}_{m}\delta^{k}_{n})
    \end{split}
\end{align}
\end{subequations}
Using \eqref{eq:Pderivativea} and \eqref{eq:PderivativeLambda}, we arrive at an expression for the subleading correction to the roots of the cubic equation
\begin{align} \label{eq:Lambdarootsol}
    \Lambda^{(1)}_{k} =
    \frac{2(\Lambda^{(0)}_k + Q_1)(\Lambda^{(0)}_k+ J_1)}{\frac{1}{2}\sum_{m \neq n \neq k}(\Lambda^{(0)}_{k} -\Lambda^{(0)}_{m})(\Lambda^{(0)}_{k}-\Lambda^{(0)}_{n})}.
\end{align}
This solution is only valid when there are three distinct roots. In the case of degenerate roots, the corrections are modified and the analysis must be done carefully. Since the coefficients of the cubic equation for $\Lambda$ are real, we expect two cases for the types of roots we may encounter: A) one real root and two complex conjugated roots or B) all real roots.\footnote{The regime where the roots are all real are defined by a constraint on the discriminant of \eqref{eq:cubicBH}: $\operatorname{disc}=-4 a_3 a_1^3+a_2^2 a_1^2+18 a_0 a_2 a_3 a_1-a_0 \left(4 a_2^3+27 a_0 a_3^2\right)>0$.}
 
 \begin{figure}
     \centering
     \includegraphics[width=\textwidth
]{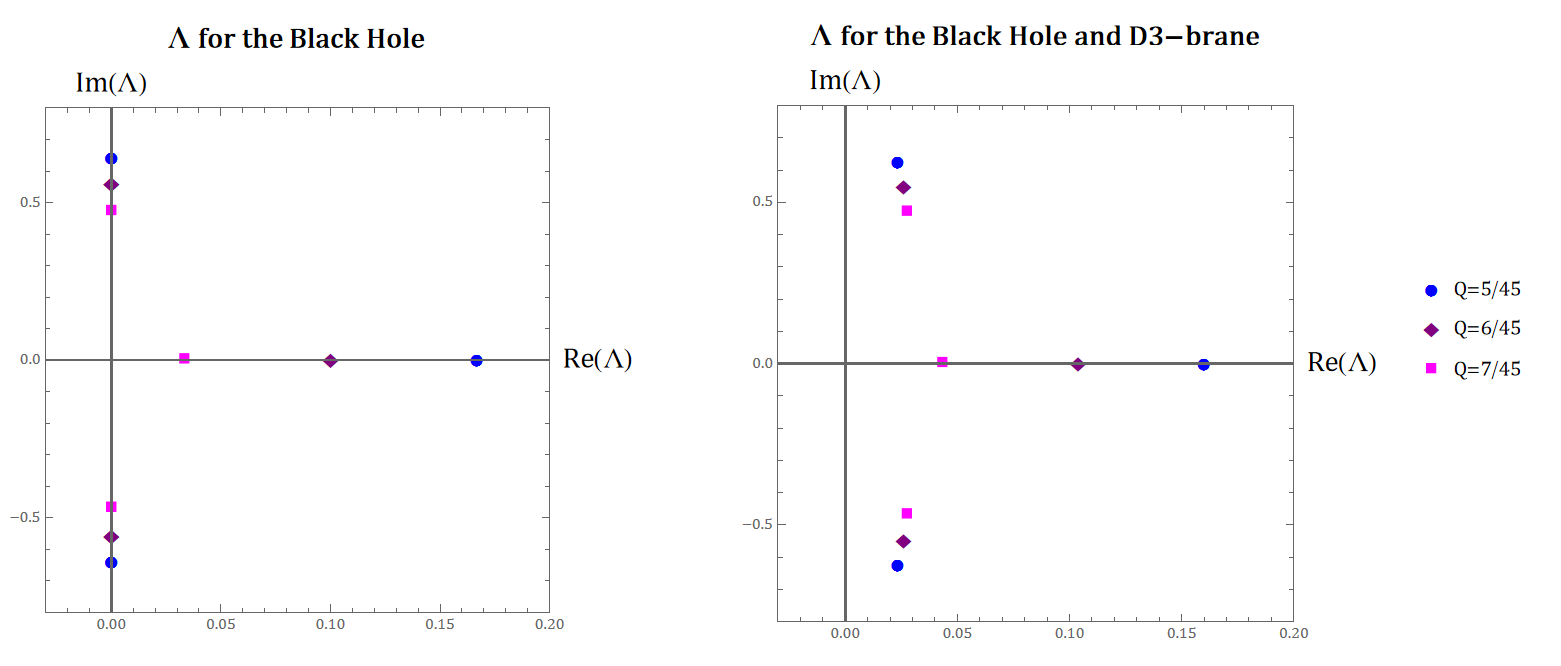}
     \caption{Here we show the complex $\Lambda$ plane with several values of two different sets of roots of \eqref{eq:cubiceq} for $N =20$, $Q_1 =Q_2 =Q_3 =Q$ and $J_1=J_2=J$ satisfying the non-linear constraint that ensures a real entropy in the pure black hole case. The first set of roots corresponds to the black hole without D$3$-brane, whereas the second corresponds to the combined system. The nonlinear constraint no longer remedies the two complex roots of $\Lambda$ to be purely imaginary when a probe D$3$-brane is introduced in the system.}
     \label{fig:Lambdaplot}
 \end{figure}

We shall focus on case A first and for simplicity, we choose a regime of charges, where all angular momenta are equal denoted by $J$ and likewise all electric charges are equal denoted by $Q$. We now revisit the problem of the nonlinear constraint
    \begin{align} \label{eq:nonlinearconstraint}
    \left(6Q+1\right)\left(3Q^2-J\right) = \left(2Q^3+J^2\right),
\end{align}
in the case of no probe brane. Then we find that the roots $\Lambda_k$ are given by
\begin{subequations}
\begin{align}
    \Lambda^{(0)}_{\pm} & =\pm i \sqrt{3 Q^2+\frac{1}{2} \left(-6 Q+\sqrt{(1-4 Q)^3}+1\right)} \, ,\quad Q<\frac{2}{9},\\
    \Lambda^{(0)}_0 & = \frac{1}{2}- 3 Q\,,
\end{align}
\end{subequations}
where $\Lambda^{(0)}_{\pm}$ are complex conjugated to each other and $\Lambda^{(0)}_0$ is real. We plot the functions of these roots on the left hand side of Figure~\ref{fig:Lambdaplot}. With the probe brane extended in the bulk, we impose yet again the nonlinear constraint \eqref{eq:nonlinearconstraint} with the charges promoted to the total charge of the combined system. We then see that the two purely imaginary roots pick up a real part and therefore get shifted while the real root also takes a smaller value, as shown in the right plot of Figure~\ref{fig:Lambdaplot}.

The key observation here is that in contradistinction to the unperturbed black holes, there is no generalization of the nonlinear constraint enforcing the reality of the entropy when the D$3$-brane is introduced. Instead, the reality of the entropy comes from the addition of the two leading gravitational saddles corresponding to the constraints~$n_0=1$ and~$n_0=-1$.

Therefore, in the case of not imposing \eqref{eq:nonlinearconstraint}, we find two of the roots take the general form
\begin{align} \label{eq:Lambdagenexpanded}
	\Lambda_{\pm} = \Lambda_{x} \pm i \Lambda_{y},
\end{align}
where $\Lambda_{x}$ and $\Lambda_{y}$ are real and can be found by taking the real and imaginary parts of \eqref{eq:Lambdarootsol}. The entropy of the system can be found by considering the sum of the two gravitational saddles, where the first saddle corresponds to the constraint $n_0=1$ for the choice~$\Lambda_{-}$ and the second saddle corresponds to~$n_0=-1$ for the choice~$\Lambda_{+}$
\begin{align}
	\begin{split}
		e^{S} &\sim
		e^{2\pi i\Lambda_-} + e^{-2\pi i \Lambda_+}
		\sim
		e^{2\pi i (\Lambda_x-i\Lambda_y)} + e^{-2\pi (\Lambda_x+i\Lambda_y)}
		=
		2 e^{2\pi \Lambda_y} \cos (2\pi \Lambda_x)\,\sim\,e^{2\pi \Lambda_y}\,,
	\end{split}
\end{align}
where
\begin{align}
    \Lambda_{y} = \frac{\gamma \left(\sqrt{3} \xi \left(3 \delta^2 (\gamma-3)+\gamma-1\right)+9 \delta^3 (3\gamma-1)\right)}{6 \left(\xi \left(54 \delta^4+18 \delta^2+1\right)+6 \sqrt{3} \left(27 \delta^2+2\right) \delta^3\right)}
\end{align}
and
\begin{align}
    \gamma^3 = 6 \delta^2 \left(\sqrt{3} \xi \delta +9 \delta ^2+3\right)+1, \quad \delta^2 = J+Q, \quad \xi^2 = 27 J+27 Q+2.
\end{align}
We stress that this result is not assuming a non-linear constraint amongst charges and we have assumed equal charges for simplicity.

Finally, we make a brief comment about scenario B. In this case, there is no predicted growth of states as the extremized value of the entropy function is purely imaginary. It would be interesting to study what happens to the growth of BPS states in this regime of charges realized in the field theory dual.

%%%%%%%%%%%%%%%%%%%%%%%%%%%%%%%%%%%%%%%%%%%%%%%%%%
%%%%%%%%%%%%%%%%%%%%%%%%%%%%%%%%%%%%%%%%%%%%%%%%%%
\section{The 4d superconformal index: A brief review}\label{section:4dindex}
%%%%%%%%%%%%%%%%%%%%%%%%%%%%%%%%%%%%%%%%%%%%%%%%%%
%%%%%%%%%%%%%%%%%%%%%%%%%%%%%%%%%%%%%%%%%%%%%%%%%%
In this section we focus on the undeformed 4d superconformal index and its integral representation. We also review the Laplace transform procedure that extracts state-degeneracies at large charges and finite values~$N$. 

The superconformal index counts (with sign) BPS states that can not combine to form long representations of the superconformal algebra. For $\mathcal{N} =1$ theories on $S^1 \times S^3$, the superconformal index was defined in \cite{Romelsberger:2007ec, Kinney:2005ej} and takes the form 
\begin{equation}
\indfd(\omega; \xi) = \text{Tr}_{\mathcal{H}(S^1\times S^3)}\left[\left(-1\right)^{F}e^{- \beta\{\mathcal{Q}, \mathcal{Q}^{\dagger}\}}e^{2 \pi i \xi_a Q_a}e^{2 \pi i \wO (J_1 + \fft{r}{2})} e^{2 \pi i \wT (J_2 + \fft{r}{2})}\right], \label{Eq:TheSCI}
\end{equation}
where $Q_a$ are the flavor charges with chemical potentials given by $\xi_a$ that will be later traded by $ \Delta_a  \equiv \xi_a +\frac{1}{2}r_a (\wO +\wT)$, where $r_a$ is the $R$-charge. The combination $J_{1,2}+\frac{r}{2}$, where $J_{1,2}$ are the angular momenta on $S^3$ and $r$ is the R-charge, commute with the supercharge $\mathcal{Q}$. The chemical potentials $\wO$ and $\wT$ are associated to $J_{1}+\frac{r}{2}$ and $J_{2}+\frac{r}{2}$, respectively. 

In particular, the superconformal index $\indfd(\wO, \wT ; \Delta)$ counts $\fft{1}{16}-$BPS states for $\mathcal{N} =4$ SYM theory and we shall focus on this theory from now on. The matter content is given by the three chiral fields $\Phi_{1,2,3}$ appearing in the superpotential
\begin{eqnarray}
W = \text{Tr} \left(\Phi_{1}\left[\Phi_{2},\Phi_3\right] \right),
\end{eqnarray}
with the associated chemical potentials being $\Delta_{1,2,3}$. For an $SU(N)$ gauge group,  $\indfd(\wO, \wT; \Delta)$ can be written as a multidimensional contour integral over the gauge holonomies  $u_{ij} \equiv u_i - u_j$ that imposes the gauge singlet constraint
\begin{eqnarray} \label{eq:Index}
\indfd \left(\wO, \wT; \Delta\right) & = & \int_{\text{SU(N)}}[\mathcal{D} U] \mathcal{Z}_{4d}(u, \wO, \wT; \Delta) \\ \nonumber
&= &\kappa_N\int_0^1\prod_{k =1}^{N-1}d u_{k}\frac{\prod_{a=1}^{3}\prod_{i \neq j}\widetilde{\Gamma} \left(u_{ij}+\Delta_a;\wO, \wT\right)}{\prod_{i \neq j}\widetilde{\Gamma} \left(u_{ij}; \wO, \wT\right)}, \label{Eq:indexN4}
\end{eqnarray}
where
\begin{align}
    \kappa_{N} &=\frac{\left(e^{2 \pi i \wO} ; e^{2 \pi i \wO} \right)_{\infty}^{N-1}\left(e^{2 \pi i \wT} ; e^{2 \pi i \wT} \right)_{\infty}^{N-1}}{(N-1) !}\prod_{a=1}^{3}\left(\widetilde{\Gamma}(\Delta_a; \wO,\wT)\right)^{N-1}.
\end{align}
We have used a modified version of the elliptic gamma function $\widetilde{\Gamma}(u;\wT, \wO)$, as described in Appendix~\ref{App:elliptic:functions}. The evaluation of \eqref{eq:Index} has been the subject of several works \cite{Choi:2018hmj, Cabo-Bizet:2018ehj, Benini:2018ywd, ArabiArdehali:2019tdm,Honda:2019cio,Cabo-Bizet:2019osg,Kim:2019yrz,Cabo-Bizet:2019eaf,Amariti:2019mgp,GonzalezLezcano2019, Lanir:2019abx,ArabiArdehali:2019orz, Murthy:2020rbd,PhysRevD.105.L021903,Agarwal:2020zwm, Benini:2020gjh,Cabo-Bizet:2020nkr,  Cabo-Bizet:2021plf,Cassani:2021fyv,Aharony:2021zkr, Cabo-Bizet:2021jar,Choi:2021rxi,Choi:2023tiq, GonzalezLezcano:2020yeb,Lezcano:2021qbj, Benini:2021ano}.   The final outcome is of the following form
\begin{align} \label{eq:Index4d1}
     \indfd(\wO,\wT; \Delta) & = N \exp \left[ - (N^2-1)\frac{i \pi}{\wO \wT}\prod_{I=1}^3
 \left( \left\{\Delta_I \right\}_{\omega}- \frac{1 - n_0}{2} \right) \right]
 , \,
 \end{align}
 where we set $ \omega= s_1^{-1} \wO = s_2^{-1} \wT  $ for some coprime integers $s_1, s_2$ (a more generic relation ). We have ignored exponentially suppressed corrections in $1/|\omega|$. The function $\{\cdot\}_{\omega}$ is defined in \eqref{tau-modded} and the value of $n_0 =\pm1$ indicates the domain of chemical potentials
 \begin{subequations}
\begin{align}
 \text{Im}\left(-\frac{1}{\omega}\right) & >\text{Im}\left(\frac{\Delta}{\omega}\right)>0\label{const1}, \quad \quad \quad & n_0 &= 1, \\
 \text{Im}\left(-\frac{1}{\omega}\right) & < \text{Im}\left(\frac{\Delta}{\omega}\right)<0\label{const2}, \quad \quad \quad & n_0 &= -1.
\end{align}
\end{subequations}
Moreover, the chemical potentials satisfy the constraint
\begin{align} \label{eq:linearconstraint1FT}
    \sum_{I=1}^3 \{ \Delta_I\}_{\omega} = \wO + \wT +\frac{3 - n_0}{2}.
\end{align}
In case of real chemical potentials with $|\Delta_{I}|<1$, for $I=1,2,3$, the leading contribution obtained in the large $N$ limit of \eqref{eq:Index4d1} gives
\begin{align}
\begin{split} 
	 \indfd(\wO,\wT; \Delta)& =N  \exp \left[-\fft{\pi i(N^2-1)}{\wO\wT}\Delta_1 \Delta_2 \Delta_3\right] \label{eq:logI10} \,,
\end{split}
\end{align}
where the linear constraint \eqref{eq:linearconstraint1FT} simplifies to
\begin{align}
     \sum_{I=1}^3 \Delta_I - \wT - \wO & = -n_0. \label{eq:constraint0}
\end{align}
As we see, the 4d index crucially depends on the domain of chemical potentials, i.e., the value of $n_0$.

%%%%%%%%%%%%%%%%%%%%%%%%%%%%%%%%%%%%%%%%%%%%%%%%%%
%%%%%%%%%%%%%%%%%%%%%%%%%%%%%%%%%%%%%%%%%%%%%%%%%%
\subsection{Extracting degeneracies: changing ensemble} \label{sec:CFTdegeneracies4d}
%%%%%%%%%%%%%%%%%%%%%%%%%%%%%%%%%%%%%%%%%%%%%%%%%%
%%%%%%%%%%%%%%%%%%%%%%%%%%%%%%%%%%%%%%%%%%%%%%%%%%
Now we would like to extract the degeneracies of the $\frac{1}{16}$-BPS states counted by the superconformal index, see for example \cite{Agarwal:2020zwm, Cabo-Bizet:2020ewf, Beccaria:2023hip}. This means that we have to change \eqref{eq:Index4d1} from the grand canonical ensemble --- with fixed chemical potentials -- to the microcanonical ensemble.
To do so, we are instructed to perform the following Laplace transformation
\begin{align} \label{eq:CFTdegeneracies4d}
   d(J; Q) & = \int d \Delta d \wT d \wO e^{- \log \indfd - 2 \pi i \sum_{I=1}^3 \Delta_I Q_I- 2 \pi i (\wO J_1+ \wT J_2) + 2 \pi i \Lambda \left(   \sum_{I=1}^3 \Delta_I - \wT -\wO + n_0\right)}\, .
\end{align}
This integration can be approximately solved in the large $N$ limit or in the Cardy-like limit ($|\wT|\, , |\wO| \ll 1$)\footnote{See \cite{ArabiArdehali:2019tdm,Honda:2019cio,Cabo-Bizet:2019osg,Kim:2019yrz,Cabo-Bizet:2019eaf,Amariti:2019mgp,Amariti:2021ubd} for extensive work on the Cardy-like limit of the superconformal index.} using the saddle point approach. We then must find the extrema of the exponent and sum over the saddle points that dominate.
\begin{figure}
 \centering
 \scalebox{0.70}{
 \begin{tikzpicture}
  \filldraw[red] (2,0) circle (2pt) node[anchor=north] {$\hspace{3.5mm} 1$}; 
 \filldraw[red] (0,0) circle (2pt) node[anchor=north] {$\hspace{3.5mm} 0$}; 
\filldraw[red] (-2,0) circle (2pt) node[anchor=north] {$-1 \hspace{8.5mm}$};
\filldraw[gray] (-4,0) circle (2pt) node[] {};
\filldraw[red] (1,2) circle (2pt) node[anchor=west] {$\hspace{1.5mm} \omega$};
\filldraw[red] (-1,2) circle (2pt) node[anchor=east] {$\omega-1\hspace{1.5mm}$};
\filldraw[red] (-1+4,2) circle (2pt) node[anchor=west] {$\hspace{1.5mm}\omega+1$};
%%%%%%%%%%%%%%%%%%%%%%%%%%%%%%%%%%%%%%%%%%%%%
\draw[->, line width = 0.6 mm, blue] (-2,0) -- (-1,0);
\draw[line width = 0.6 mm, blue] (-1.2,0) -- (0,0);
\draw[->,line width = 0.6 mm, blue] (0,0) -- (1,0);
\draw[line width = 0.6 mm, blue] (0.9,0) -- (2,0);
%%%%%%%%%%%%%%%%%%%%%%%%%%%%%%%%%%%%%%%%%%%%%%%%%%
\draw[dashed,thick] (-2,0) parabola[parabola height=1.0cm] +(2,0);
\filldraw[purple] (-1,1) circle (2pt) node[anchor=south] {$\vspace{1.5mm}\Delta_{+}$};
\draw[dashed,thick] (0,0) parabola[parabola height=-1.0cm] +(2,0);
\filldraw[purple] (1,-1) circle (2pt) node[anchor=north] {$\vspace{1.5mm}\Delta_{-}$};
%%%%%%%%%%%%%%%%%%%%%%%%%%%%%%%%%%%%%%%%%%%%%%%%%%%
\draw[->,thick, black] (-4,0) -- (4,0);
\draw[->,thick, black] (0,-2)--(0,5);
\draw[thick,blue!80!red] (-3,-2) -- (0.5,5);
\draw[thick,blue!80!red] (-1,-2) -- (2.5,5);
\draw[dashed] (1,-2) -- (4.5,5);
\draw[dashed] (-5,-2) -- (-1.5,5);
\draw[fill=gray!50, nearly transparent]  (-3,-2) -- (0.5,5) -- (2.5,5) -- (-1,-2) -- cycle;
\draw[fill=blue!50, nearly transparent]   (-3+2,-2) -- (0.5+2,5) -- (2.5+2,5) -- (-1+2,-2) -- cycle;
 %%%%%%%%%%%%%%%%%%%%%%%%%%%%%%%%%%%%%%%%%%%%%%%%%%%%%%%
 %%%%%%%%%%%%%%%%%%%%%%%%%%%%%%%%%%%%%%%%%%%%%%%%%%%%%%
 %%%%%%%%%%%%%%%%%%%%%%%%%%%%%%%%%%%%%%%%%%%%%%%%%%%%%%%

\end{tikzpicture}
}
\caption{The Figure shows the complex plane of chemical potentials for a generic $\Delta$ where the region corresponding to $n_0=1$ (\ref{const1}) is shown in gray and the region specified by $n_0=-1$ (\ref{const2}) is shown in light blue. The thick blue arrows running over the interval $\text{Re}(\Delta) \in (-1;\, 1)$ represent the integration contour for $\Delta_I$. With dashed curves we schematically show a deformation of the initial contour such that it passes through the critical values of chemical potentials $\Delta_{\pm}$.} \label{fig:DeltaDomain1}
\end{figure}
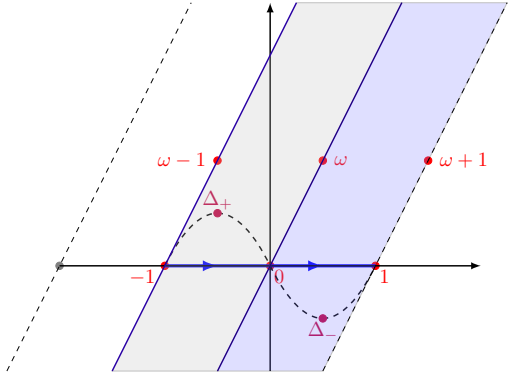
Let us consider a family of critical points $\{\Delta_{n_0} ,\, \omega_{n_0}\}$, as shown schematically in Figure~\ref{fig:DeltaDomain1}, such that the effective action has the same real part when evaluated at these points. In other words, both saddles are equally leading in the saddle point approximation which then gives us
\begin{align}
    d(J; \, Q)  & \approx \sum_{n_0= \pm 1 } \left( e^{- \log \indfd - 2 \pi i \sum_{I=1}^3 \Delta_I Q_I- 2 \pi i(\wO J_1+ \wT J_2) + 2 \pi i \Lambda \left(   \sum_{I=1}^3 \Delta_I - \wT -\wO + n_0\right)}\right)\Big{|}_{\Delta_{n_0}} \,.
\end{align}
Note that here we identify the critical points using only $\Delta_{n_0}$ because the constraint \eqref{eq:constraint0} already determines the critical values of $\omega_{n_0}$. The extremization leads us to the following relations
\begin{subequations}
\begin{align} \label{eq:SPlogI}
    \frac{\partial \log \mathcal{I}_{4d}}{\partial \Delta_I} &=   2 \pi i (\Lambda- Q_I  ), \, \quad I=1, \cdots 3\, , \\
    \frac{\partial \log \mathcal{I}_{4d}}{\partial \wO} &=   2 \pi i (J_1 +  \Lambda),  \, \\
     \frac{\partial \log \mathcal{I}_{4d}}{\partial \wT} &= 2 \pi i (J_2 +  \Lambda),
\end{align}
\end{subequations}
under the constraint \eqref{eq:constraint0}. This implies
 \begin{align} \label{eq:cubic1}
  2  \prod_{I=1}^3 (Q_I  - \Lambda) & =  N^2 (J_1 +\Lambda)(J_2 +\Lambda) .
 \end{align}
 The relation \eqref{eq:cubic1}  have precisely the same structure as \eqref{eq:cubiceq}, without the last term that accounts for the D$3$-brane contributions and assuming the charges are the ones of the 4d $\frac{1}{16}$-BPS states counted by the superconformal index. The degeneracy of states is then given by
\begin{align} \label{eq:Sn0}
    d(J; \, Q)  & \,\sim\, \sum_{n_0= \pm 1 }  \sum_{k =\pm, 0} e^{-2\pi i n_0 \Lambda_k} \,.
\end{align}

 %\subsection{The cubic equation}

Generically, if we require a growth of states, then \eqref{eq:cubic1} must have two complex conjugated solutions for $\Lambda$, which we have called $\Lambda_{\pm}$ in \eqref{eq:Lambdagenexpanded}, that upon application of \eqref{eq:Sn0} generate a dominant saddle for each value of $n_0$. These dominant saddles correspond to the roots satisfying
$\text{Re}(2 \pi i n_0 \Lambda_{\pm}) >0$. If we appropriately label $\Lambda_{\pm}$ such that the subindex corresponds to the sign of its imaginary part, we have
 \begin{align}
     d(J;\, Q) & \sim e^{-2 \pi i \Lambda_+} + e^{2 \pi i \Lambda_-} = 2e^{2 \pi \Lambda_y} \cos\left(2 \pi \Lambda_x\right)\,\sim\, e^{2 \pi \Lambda_y}.
     \end{align}
If the imaginary part of $\Lambda_{\pm}$ vanishes, then we see that the microcanonical expression for the index is a pure oscillatory term that does not probe the growth of states compatible with black hole entropy.  

\section{The defect superconformal index} \label{sec:defectIndex}
The computation of the defect index requires the consistent embedding of the 2d $\mathcal{N}=(2,2)$ superconformal algebra into the 4d $\mathcal{N}=4$ superconformal algebra. This has been done in detail for $\mathcal{N}=2$ \cite{Gadde:2013dda} as well as for $\mathcal{N}=4$ \cite{Chen:2023lzq}. In these works, the fugacities used in the 2d description are related to those in 4d. For this reason, we write the full defect index purely in terms of the 4d chemical potentials. 

The defect worldvolume $\mathcal{M}_{2d}$ extends along the $S^1$ and wraps a circle inside the $S^3$. The surface operator with support on this $\mathcal{M}_{2d}$ will be such that it commutes with the supercharge selected to construct the 4d superconformal index, and this is ensured in practice by appropriately choosing the orientation of the surface operator during the embedding procedure. The defect index is then given by
    \begin{align} \label{eq:ID}
        \mathcal{I}_{\text{D}} & = \int_{\text{SU(N)}}[\mathcal{D} U] \mathcal{Z}_{4d}(u, \wO, \wT; \Delta) \mathcal{Z}_{2d}(u, \wO, \wT; \Delta) ~.
    \end{align}
It has been proposed that \eqref{eq:ID} provides the microscopic
definition of a dual gravity system which includes black holes interacting with a probe D$3$-brane \cite{Chen:2023lzq}. We revisit this matter carefully in this section.

It is possible to work in the approximation where the saddles of \eqref{eq:Index} are not affected by the insertion of $\mathcal{Z}_{2d}$ in \eqref{eq:ID}. This regime corresponds holographically to the probe limit of the black hole/D$3$-brane system. In this probe limit we can write
 \begin{align} \label{eq:4d2dProbe}
     \mathcal{I}_{\text{D}} & = \sum_{\hat{u} \in 4d-\text{saddles}} \mathcal{Z}_{4d}(\hat{u}, \wO, \wT; \Delta) \mathcal{Z}_{2d}(\hat{u}
    , \wO, \wT; \Delta)  + \cdots \,,
 \end{align}
where the $\cdots$ correspond to the subleading saddles. Instead of directly implementing the probe limit \eqref{eq:4d2dProbe}, we first study the 2d index in the context of a systematic Cardy-like expansion along the lines of \cite{GonzalezLezcano:2020yeb, GonzalezLezcano:2022hcf}. This enables us to have better control over the effect of backreaction coming from the 2d defect on the 4d index. If we denote the fundamental domains of chemical potentials $\Delta^{(n_0)}$, $n_0 = \pm1$ then we will see in Subsection~\ref{section:2dindex} that for $n_0=1$, the integrand $\mathcal{Z}_{2d}$ in \eqref{eq:ID} becomes independent of the holonomies up to corrections exponentially suppressed in $1/|\omega|$
 \begin{align} \label{eq:IDm1}
      \mathcal{I}^{(1)}_{\text{D}} & =    \mathcal{Z}_{2d}( \wO, \wT; \Delta^{(1)})\sum_{\hat{u} \in 4d-\text{saddles}} \mathcal{Z}_{4d}(\hat{u}, \wO, \wT; \Delta^{(1)})\, .
 \end{align}
Moreover, for the other domain of chemical potentials labelled by $n_0=-1$, the leading order in the Cardy-like limit is given by
 \begin{align} \label{eq:IDp1}
      \mathcal{I}^{(-1)}_{\text{D}}\Big{|}_{\omega\rightarrow 0} & =    \mathcal{Z}_{2d}( \wO, \wT; \Delta^{(-1)})\sum_{\hat{u} \in 4d-\text{saddles}} \mathcal{Z}_{4d}(\hat{u}, \wO, \wT; \Delta^{(-1)})\, .
 \end{align}
We now turn to study the 2d index in the systematic Cardy-like expansion.
 
\subsection{The Cardy-like expansion of the 2d index} \label{section:2dindex}

Following \cite{Chen:2023lzq, Gadde:2013dda}, we start with the 2d index given as
\begin{eqnarray}
    \mathcal{Z}_{2d} & = & \sum_{i=1}^N \exp \Biggl[  \sum_{j \neq i} \log  \frac{\theta_0(  -u_{ij} - \Delta_2 + \wO; \, \wO)}{\theta_0( -u_{ij} + \Delta_1 -  \wT; \, \wO)} + \log \frac{\theta_0( u_{ij} - \Delta_1- \Delta_2 +\wT+ \wO; \, \wO)}{\theta_0( u_{ij}  ; \, \wO)} \Biggr].
\end{eqnarray}
Now using the elliptic theta functions in \eqref{eq:theta:0} and \eqref{eq:theta:01}, the 2d index can be recast in the form
\begin{align}
\begin{split}
    \mathcal{Z}_{2d} & = \sum_{i =1}^N \exp \left[  \sum_{j \neq i} \log \frac{\left(\, {\rm e}^{ 2 \pi \i ( -u_{ij} - \Delta_2 + \wO)} ; \, {\rm e}^{2 \pi \i \wO}\right)_\infty \left(\, {\rm e}^{ 2 \pi \i ( u_{ij} + \Delta_2 )} ; \, {\rm e}^{2 \pi \i \wO}\right)_\infty}{\left(\, {\rm e}^{ 2 \pi \i (-u_{ij} + \Delta_1 -  \wT)} ; \, {\rm e}^{2 \pi \i \wO}\right)_\infty \left(\, {\rm e}^{ 2 \pi \i ( u_{ij} - \Delta_1 +\wT + \wO )} ; \, {\rm e}^{2 \pi \i \wO}\right)_\infty} \right.  \\
    & \left. \quad \quad +  \sum_{j \neq i} \log \frac{\left(\, {\rm e}^{ 2 \pi \i ( u_{ij}- \Delta_1 - \Delta_2+ \wT  + \wO)} ; \, {\rm e}^{2 \pi \i \wO}\right)_\infty \left(\, {\rm e}^{ 2 \pi \i ( -u_{ij} +\Delta_1+ \Delta_2 + \wT)} ; \, {\rm e}^{2 \pi \i \wO}\right)_\infty}{\left(\, {\rm e}^{ 2 \pi \i ( u_{ij} )} ; \, {\rm e}^{2 \pi \i \wO}\right)_\infty \left(\, {\rm e}^{ 2 \pi \i ( \wO -u_{ij})} ; \, {\rm e}^{2 \pi \i \wO}\right)_\infty} \right].
    \end{split}
\end{align}
We consider $u_{ij} = (x_{ij} \wO + y_{ij} \wT$) and $\wT = \frac{s_1}{s_2} \wO$ such that $u_{ij} = \wO (x_{ij} +  \frac{s_1}{s_2} y_{ij}) \equiv \wO z_{ij}$. These change of variables allows us to implement the systematic Cardy-like expansion.\footnote{See \cite{GonzalezLezcano:2022hcf} for a similar implementation to the 3d superconformal index.} Upon using the asymptotic expansion \eqref{Eq:Pochhamer-IdentityMain}, we find
\begin{align}
\begin{split}
    \mathcal{Z}_{2d} & = \sum_{i =1}^N \exp \left[ \frac{1}{2 \pi \i \wO}  \sum_{j \neq i}\sum_{r =0}^{\infty} (-1)^r \frac{(2 \pi \i \wO)^r}{r !} \left( B_r (z_{ij}) \text{Li}_{2 - r} \left({\rm e }^{2 \pi \i ( - \Delta_2)}\right)  \right. \right. \\
    & \quad \quad \quad \quad \left. \left. + B_r \left(1-z_{ij}\right) \text{Li}_{2 - r} \left( { \rm e}^{2 \pi \i( \Delta_2)}\right) + B_r \left(-\frac{s_1}{s_2}- z_{ij}\right) \text{Li}_{2 - r} \left({\rm e}^{ 2 \pi \i (- \Delta_1 - \Delta_2 )} \right)  \right. \right. \\
    &\quad \quad \quad \quad   \left. \left.+ B_r \left(1+\frac{s_1}{s_2} +z_{ij}\right) \text{Li}_{2 - r} \left({\rm e}^{2 \pi \i( \Delta_1 +\Delta_2 )}\right)  \right. \right. \\
    &\quad \quad \quad \quad  \left. \left.- B_r \left(1+\frac{s_1}{s_2} +z_{ij}\right) \text{Li}_{2 - r} \left({\rm e}^{2\pi \i ( \Delta_1)} \right)  \right. \right. \\
    &\quad \quad \quad \quad  \left. \left.- B_r \left(-\frac{s_1}{s_2}-z_{ij}\right) \text{Li}_{2 - r} \left({\rm e}^{ 2 \pi \i ( - \Delta_1  ) }\right) \right. \right. \\ 
    &\quad \quad \quad \quad  \left. \left.- B_r \left(1- z_{ij}\right) \text{Li}_{2 - r} \left({\rm e}^{ 2 \pi \i (\epsilon) }\right)- B_r \left(z_{ij}\right) \text{Li}_{2 - r} \left({\rm e}^{ 2 \pi \i (- \epsilon) }\right)  \right)\right].
    \end{split}
\end{align}
In the last line we have regulated the Polylogarithms via a small $\epsilon >0$ regulator. We will see that we can safely take $\epsilon \rightarrow 0$ at the end of the manipulations. Utilizing the property of Bernoulli polynomials,
\begin{subequations}
\begin{align}
    B_r (1 -x) & = (-1)^r B_r (x), \, \quad r \geq 0\, , \\
 B_r(0)   & = (-1)^r B_r(1)\,,
\end{align}
\end{subequations}
the 2d index is simplified to
\begin{align}
\begin{split} \label{eq:intermedstep}
    \mathcal{Z}_{2d} & = \sum_{i =1}^N \exp \left[ \frac{1}{2 \pi \i \wO}  \sum_{j \neq i} \sum_{r =0}^{\infty}  \frac{(2 \pi \i \wO)^r}{r !}  \left[ B_r(z_{ij})\left((-1)^r \text{Li}_{2 - r} \left({\rm e }^{2 \pi \i ( - \Delta_2)}\right) + \text{Li}_{2 - r} \left( { \rm e}^{2 \pi \i(\Delta_2)}\right)\right) \right. \right. \\
    & \quad \quad \quad \quad \left. \left.+ B_r(-\frac{s_1}{s_2}- z_{ij})\left((-1)^r \text{Li}_{2 - r} \left({\rm e}^{ 2 \pi \i ( - \Delta_1 - \Delta_2 )} \right) + \text{Li}_{2 - r} \left({\rm e}^{2 \pi \i( \Delta_1 +\Delta_2 )}\right) \right) \right. \right. \\
    & \quad \quad \quad \quad \left. \left.- B_r(-\frac{s_1}{s_2}- z_{ij}) \left( \text{Li}_{2 - r} \left({\rm e}^{2\pi \i ( \Delta_1 )} \right) + (-1)^r \text{Li}_{2 - r} \left({\rm e}^{ 2 \pi \i ( - \Delta_1  ) }\right)\right) \right. \right. \\ 
    &\quad \quad \quad \quad  \left. \left.-B_r(z_{ij}) \left(\text{Li}_{2 - r} \left({\rm e}^{ 2 \pi \i (\epsilon) }\right)+ (-1)^r \text{Li}_{2 - r} \left({\rm e}^{ 2 \pi \i (- \epsilon) }\right)\right)  \right]\right]\,.
    \end{split}
\end{align}
The final simplification requires us to use the property of Polylogarithm functions
\begin{subequations}
\begin{align} \label{eq:identityLi}
    \text{Li}_n ({\rm e}^{2 \pi \i z}) + (-1)^n \text{Li}_n ({\rm e}^{-2 \pi \i z}) &=  -  \frac{(2 \pi \i)^n}{n!} B_n (\{z\}), \quad n = 1,2,3, \cdots \,, \\
      \text{Li}_{-n} ({\rm e}^{2 \pi \i z}) + (-1)^n \text{Li}_{-n} ({\rm e}^{-2 \pi \i z}) &=  0, \quad n = 0, 1,2,3, \cdots \,,
\end{align}
\end{subequations}
for $ 0 \leq\text{ Re}(z) < 1$ and $\text{Im}(z) \geq 0$ or $ 0 < \text{ Re}(z) \leq 1$ and $\text{Im}(z) < 0$. Now \eqref{eq:intermedstep} becomes
\begin{align}
\begin{split} \label{eq:I2dExact}
    \mathcal{Z}_{2d} & = \sum_{i =1}^N \exp \left[ \frac{1}{2 \pi \i \wO}  \sum_{j \neq i}\sum_{r =0}^{1}  \frac{(2 \pi \i)^{2-r}}{(2-r)!}\frac{(2 \pi \i \wO)^r}{r!}  \left( - B_{r} (z_{ij}) \left(B_{2-r}\left(\{ \Delta_2\}\right) -  B_{2 - r} \right) \right. \right. \\
    & \left. \left.  -  (-1)^{2- r} B_r(-\frac{s_1}{s_2}-z_{ij}) \left( B_{2 - r} \left(\{ - \Delta_1 - \Delta_2 \} \right)  -  B_{2-r} \left(\{ - \Delta_1 \} \right) \right) +   \right)\right] \\
    & = \sum_{i =1}^N \exp \left\{ \frac{2 \pi \i (N-1) }{ \wO} \left[ \prod_{a=2}^3\left(\{\Delta_a\} - n\right)   +  \delta_{n_0,- 1}\left(\sum_{j\neq i}  \frac{u_{ij}}{N-1} -\frac{\wO}{2} \right)\right]\right\} \,,
    \end{split}
\end{align}
where we have recovered the original holonomy variables, namely $u_{ij} = \frac{ z_{ij}}{\wO}$ and for compactness we have defined $n \equiv \frac{1-n_0 }{2}$. 
The function $\{z\}$ defined in \eqref{modded} is such that it forces the Bernoulli polynomials to have the same periodicity properties as the Polylogarithm functions. Note that \eqref{eq:identityLi} ensures that the terms with the $\epsilon$ regulator produce a finite result as the right hand side is a finite quantity at $z =0$. 

For the leading correction in the Cardy-like limit, the last terms in \eqref{eq:I2dExact} vanishes and we have
\begin{align} \label{eq:fomega}
    \mathcal{Z}_{2d} =N \exp \left[\frac{2 \pi \i (N-1) }{ \wO} \left( \prod_{a=2}^3\left(\{\Delta_a\} -  n\right)\right)\right] \, .
\end{align}
There are two main observations to make regarding the transition from \eqref{eq:I2dExact} to \eqref{eq:fomega}:

\begin{itemize}
    \item For the fundamental domain of chemical potentials labelled by $n_0 = 1$, the holonomies drop from the expression \eqref{eq:I2dExact}, rendering our result \eqref{eq:fomega} valid at finite $N$ up to exponentially supressed corrections in $1/|\omega|$. This domain of chemical potentials corresponds to \eqref{eq:IDm1} and  clearly in this case there is no need to work in the probe limit. 

    \item For $n_0 =-1$  there is a linear dependence of the holonomies in \eqref{eq:I2dExact} that accounts for backreaction of the D$3$-brane when considering the combined 4d-2d system. However, this term vanishes in the leading term of the Cardy-like limit. This allow us to factor out the 2d index contribution to the 4d defect superconformal index, as anticipated in \eqref{eq:IDp1}.
\end{itemize}

The next step is to explicitly evaluate \eqref{eq:ID} and extract the microcanonical degeneracies implementing the Laplace transformation.

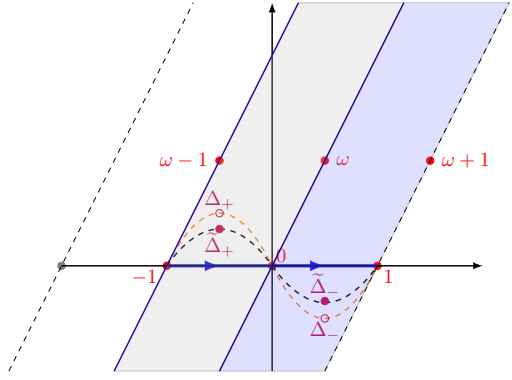
\begin{figure}
 \centering
 \scalebox{0.70}{
 \begin{tikzpicture}
  \filldraw[red] (2,0) circle (2pt) node[anchor=north] {$\hspace{4.2mm} 1$}; 
 \filldraw[red] (0,0) circle (2pt) node[anchor=south] {$\hspace{3.5mm} 0$}; 
\filldraw[red] (-2,0) circle (2pt) node[anchor=north] {$-1 \hspace{8.5mm}$};
\filldraw[gray] (-4,0) circle (2pt) node[] {};
\filldraw[red] (1,2) circle (2pt) node[anchor=west] {$\hspace{1.5mm} \omega$};
\filldraw[red] (-1,2) circle (2pt) node[anchor=east] {$\omega-1\hspace{1.5mm}$};
\filldraw[red] (-1+4,2) circle (2pt) node[anchor=west] {$\hspace{1.5mm}\omega+1$};
%%%%%%%%%%%%%%%%%%%%%%%%%%%%%%%%%%%%%%%%%%%%%
\draw[->, line width = 0.6 mm, blue] (-2,0) -- (-1,0);
\draw[line width = 0.6 mm, blue] (-1.2,0) -- (0,0);
\draw[->,line width = 0.6 mm, blue] (0,0) -- (1,0);
\draw[line width = 0.6 mm, blue] (0.9,0) -- (2,0);
%%%%%%%%%%%%%%%%%%%%%%%%%%%%%%%%%%%%%%%%%%%%%%%%%%
\draw[orange,dashed,thick] (-2,0) parabola[parabola height=1.0cm] +(2,0);
\draw[purple] (-1,1) circle (2pt) node[anchor=south] {$\vspace{1.5mm}\Delta_{+}$};
\draw[orange,dashed,thick] (0,0) parabola[parabola height=-1.0cm] +(2,0);
\draw[purple] (1,-1) circle (2pt) node[anchor=north] {$\vspace{1.5mm}\Delta_{-}$};

%%%%%%%%%%%%%%%%%%%%%%%%%%%%%%%%%%%%%%%%%%%%%%%%%%%%
\draw[dashed,thick] (-2,0) parabola[parabola height=0.7cm] +(2,0);
\filldraw[purple] (-1,0.7) circle (2pt) node[anchor=north] {$\vspace{1.5mm}\widetilde{\Delta}_{+}$};
\draw[dashed,thick] (0,0) parabola[parabola height=-0.7cm] +(2,0);
\filldraw[purple] (1,-0.67) circle (2pt) node[anchor=south] {$\vspace{2.5mm}\widetilde{\Delta}_{-}$};

%%%%%%%%%%%%%%%%%%%%%%%%%%%%%%%%%%%%%%%%%%%%%%%%%%%
\draw[->,thick, black] (-4,0) -- (4,0);
\draw[->,thick, black] (0,-2)--(0,5);
\draw[thick,blue!80!red] (-3,-2) -- (0.5,5);
\draw[thick,blue!80!red] (-1,-2) -- (2.5,5);
\draw[dashed] (1,-2) -- (4.5,5);
\draw[dashed] (-5,-2) -- (-1.5,5);
\draw[fill=gray!50, nearly transparent]  (-3,-2) -- (0.5,5) -- (2.5,5) -- (-1,-2) -- cycle;
\draw[fill=blue!50, nearly transparent]   (-3+2,-2) -- (0.5+2,5) -- (2.5+2,5) -- (-1+2,-2) -- cycle;

 %%%%%%%%%%%%%%%%%%%%%%%%%%%%%%%%%%%%%%%%%%%%%%%%%%%%%%%
 %%%%%%%%%%%%%%%%%%%%%%%%%%%%%%%%%%%%%%%%%%%%%%%%%%%%%%
 %%%%%%%%%%%%%%%%%%%%%%%%%%%%%%%%%%%%%%%%%%%%%%%%%%%%%%%

\end{tikzpicture}
}
\caption{The Figure shows the complex plane of chemical potentials for a generic $\Delta$ where the region corresponding to $n_0=1$ (\ref{const1}) is shown in gray and the region specified by $n_0=-1$ (\ref{const2}) is shown in light blue. The thick blue arrows running over the interval $\text{Re}(\Delta) \in (-1;\, 1)$ represent the integration contour for $\Delta_I$. Now we represent the deformed contour in the presence of the D$3$-brane in black dashed lines passing throught the new saddles labeled as $\widetilde{\Delta}_{\pm}$. We have kept the contour (dashed orange curve) for the case of the black hole in the absence of D$3$-brane just for reference.   }  \label{fig:Deltadeform2}
\end{figure}

%%%%%%%%%%%%%%%%%%%%%%%%%%%%%%%%%%%%%%%%%%%%%%%%%
%%%%%%%%%%%%%%%%%%%%%%%%%%%%%%%%%%%%%%%%%%%%%%%%%
\subsection{The combined 4d-2d system: extracting  degeneracies}
%%%%%%%%%%%%%%%%%%%%%%%%%%%%%%%%%%%%%%%%%%%%%%%%%
%%%%%%%%%%%%%%%%%%%%%%%%%%%%%%%%%%%%%%%%%%%%%%%%%
For simplicity, we continue to restrict ourselves to the leading term in the Cardy-like limit when we consider the combined 4d-2d system. From  \eqref{eq:Index4d1}, \eqref{eq:IDm1}, \eqref{eq:IDp1} and \eqref{eq:fomega}, the total defect index is given by
\begin{align} \label{eq:IDn}
    \mathcal{I}_{\text{D}} = 
       N \text{exp} \left[ \frac{ i \pi (N-1)}{ \wO} \left(- \frac{(N+1)}{ \wT}\prod_{I=1}^3
 \left( \left\{\Delta_I \right\}_{\omega}-n \right)  +  2 \prod_{a=2}^3\left(\{\Delta_a\} -  n\right)\right) \right] .
\end{align}
Note that for purely imaginary $\wO, \wT$ as well as for purely real arguments, the functions $\{\cdot\}_{\omega}$ and $\{\cdot\}$ coincide, which allows the expressions for $\mathcal{Z}_{2d}$ and $\mathcal{Z}_{4d}$ to be written in terms of the same combinations of chemical potentials.
The expression \eqref{eq:IDn} considerably simplifies in the regimes of real chemical potentials $\Delta_I$ such that $|\Delta_I|<1$
\begin{align} \label{eq:GCIndexD}
    \mathcal{I}_{\text{D}} & = 
       N \text{exp} \left[ \frac{ i \pi (N-1)}{ \wO} \left(- \frac{(N+1)}{ \wT}\Delta_1 \Delta_2\Delta_3  + 2 \Delta_2 \Delta_3 \right) \right] \, , \\
      \Delta_3 & = \wO  + \wT -\Delta_1 -\Delta_2  - n_0 \, .
\end{align}
From now on we proceed to extract the microcanonical degeneracies from the grandcanonical expression for the defect index given in \eqref{eq:GCIndexD}. Implementing the Laplace transform using the saddle point method requires the following extremization process
\begin{align} \label{eq:SPlogID}
    \frac{\partial \log \mathcal I_{\text{D}}}{\partial \Delta_I} &=   2 \pi i (\Lambda- Q_I  ), \, \quad I=1, \cdots 3\, , \\
    \frac{\partial \log \mathcal I_{\text{D}}}{\partial \wO} &=  - 2 \pi i (J_1 +  \Lambda)\, , \\
      \frac{\partial \log \mathcal I_{\text{D}}}{\partial \wT} &=  - 2 \pi i (J_2 +  \Lambda) \,,
\end{align}
under the constraint \eqref{eq:constraint0}. It is worth pointing out that in this case the charges $Q_{1,2,3}$ and $J_{1,2}$ are the total charge of 4d and 2d states.

From this point on, the mathematical problem is essentially equivalent to \eqref{eq:extremizationDelta}. In the field theoretical language we have to repeat the calculation of Section~\ref{sec:CFTdegeneracies4d} just replacing $\mathcal{I}_{4d}$ by $\mathcal{I}_{\text{D}}$ in \eqref{eq:CFTdegeneracies4d}. If we focus on the large $N$ regime, following the logic of Section~\ref{subsec:ExtremizationGravity}, we find a new set of saddle points through solving a modified cubic equation completely equivalent to \eqref{eq:cubiceq}. In Figure~\ref{fig:Deltadeform2}, we show a schematic picture of how the new saddles in the complex domain of chemical potentials can be changed by the $1/N$ corrections introduced by considering the contribution of the 2d defect states to the degeneracy.

%%%%%%%%%%%%%%%%%%%%%%%%%%%%%%%%%%%%%%%
%%%%%%%%%%%%%%%%%%%%%%%%%%%%%%%%%%%%%%%

%%%%%%%%%%%%%%%%%%%%%%%%%%%%%%%
%%%%%%%%%%%%%%%%%%%%%%%%%%%%%%%%%

%%%%%%%%%%%%%%%%%%%%%%%%%%%%%%
\section{Final comments and open questions} \label{section:conclusions}
%%%%%%%%%%%%%%%%%%%%%%%%%%%%%%

We have presented the thermodynamic analysis for the combined black hole/D$3$-brane system in a way that ensures the reality of the charges and the entropy. To do so, we took into account both leading saddles of the gravitational path integral and likewise, on the gauge dual, the two leading saddles of the integral over the holonomies that represent the defect superconformal index. Following this procedure we obtain real charges and entropy without the need of imposing a nonlinear constraint among the charges.

There are various interesting extensions of this work. The first is to compute the backreaction of the D$3$-brane in the geometry. One first step in this direction may be solving the Killing spinor equations for the combined system of a black hole/D$3$-brane. Presumably, while doing so, we would be forced into computing relevant backreaction effects. This may allow us to answer questions like whether the exact change in entropy predicted by the field theory side of the duality can be interpreted as a change in the area of the black hole horizon.
	
\section*{Acknowledgments}
We thank Davide Cassani, Robie Hennigar and Enrico Turetta for fruitful for discussions. MD is supported in part by the Odysseus grant (G0F9516N Odysseus) from the as well as the the Postdoctoral Fellows of the Research Foundation - Flanders grant (1235324N).
AGL is supported by an appointment to the JRG Program at the APCTP through the Science and Technology 
Promotion Fund and Lottery Fund of the Korean Government, by the Korean Local 
Governments - Gyeongsangbuk-do Province and Pohang City, and  by the National 
Research Foundation of Korea (NRF) grant funded by the Korea government (MSIT) (No. 2021R1F1A1048531).
The authors would would like to thank the Isaac Newton Institute for Mathematical Sciences, Cambridge, for support and hospitality during the programme Black holes: bridges between number theory and holographic quantum information where work on this paper was undertaken. This work was supported by EPSRC grant no EP/R014604/1.

	 \appendix

\section{Revisiting a previous approach} \label{appendix:reviewofpreviousapproach}
%%%%%%%%%%%%%%%%%%%%%%%%%%%%%%%%
%%%%%%%%%%%%%%%%%%%%%%%%%%%%%%%%

In this appendix we briefly review the procedure used in~\cite{Chen:2023lzq} to compute the entropy of the combined system. Recalling that at the level of the on-shell action 
\be
I = I_{\text{BH}}+I_{\text{D3}}
\ee
and requiring that
\be
S = S_{\text{BH}}+ S_{\text{D3}},
\ee
then the first law of thermodynamic of the total system would unequivocally constrain the thermodynamic charges of the D3-brane to be

\begin{align} \label{eq:JQdecouplinglimit}
    J_{1,\text{D3}} &= -\frac{1}{\beta} \frac{\partial I_{\text{D3}}}{\partial \Omega_{1}} 
    , \quad
    J_{2,\text{D3}} = -\frac{1}{\beta} \frac{\partial I_{\text{D3}}}{\partial \Omega_{2}} 
    , \quad
    Q_{\text{D3}} = -\frac{1}{\beta} \frac{\partial I_{\text{D3}}}{\partial \Phi}\,,
\end{align}
provided the chemical potentials do not receive subleading corrections in~$1/N$, or equivalently, that they are the very same chemical potentials of the unperturbed black hole. To evaluate the right hand sides of the equations in \eqref{eq:JQdecouplinglimit}, we must invert the Jacobian matrix
\begin{align}
    \frac{\partial(\Omega_{1}, \Omega_{2}, \Phi, \beta)}{\partial (a,b,q,r_+)}.
\end{align}
In this way we find the expressions for the electric charges and angular momentum of the D3-brane reported in~\cite{Chen:2023lzq}.

\subsection{Legendre transform with just the D3-brane}
%%%%%%%%%%%%%%%%%%%%%%%%%%%%%%%%
%%%%%%%%%%%%%%%%%%%%%%%%%%%%%%%%
The entropy of the D3-brane is defined from the Legendre transform
\begin{align}
    \begin{split}
    S_{\text{D3}} &= - I_{\text{D3}} -  2 \pi i \sum_{I=1}^3\varphi_{I,\text{BH}} Q_{I,\text{D3}} - 2\pi i \sum_{k=1}^2 \omega_{k,\text{BH}} J_{k,\text{D3}}  + 2\pi i \Lambda \left(\sum_{I=1}^3\varphi_{I,\text{BH}} - \sum_{k=1}^2\omega_{k,\text{BH}} + n_0 \right),
    \end{split}
\end{align}
where $\Lambda$ is a Lagrange multiplier implementing the corresponding linear constraint, and we have reinstated the subindex BH to recall that these are the very same potential as the black hole solution. The extremization leads to the following equations
\begin{align} \label{eq:exteqforD3}
		0&=-\frac{\partial I_{\text{D3}}}{\partial \varphi_{I, \text{BH}}} - 2\pi i (Q_{I,\text{D3}} - \Lambda) = 2\pi i \left(N \frac{\varphi_{2,\text{BH}}\varphi_{3,\text{BH}}}{\varphi_{I, \text{BH}} \omega_{1,\text{BH}}}(\delta^I_2+\delta^I_3) - Q_{I,\text{D3}} + \Lambda \right), & I&=1,2,3, \nonumber
		\\
		0&=-\frac{\partial I_{\text{D3}}}{\partial \omega_{k, \text{BH}}} - 2\pi i (J_{k,\text{D3}} + \Lambda) = - 2\pi i \left(N \frac{\varphi_{2, \text{BH}}\varphi_{3, \text{BH}}}{\omega_{k, \text{BH}} \omega_{1, \text{BH}}}\delta^k_1 + J_{k,\text{D3}} + \Lambda \right), & k&=1,2.
\end{align}

Imposing \eqref{eq:exteqforD3}, we find that the entropy is given by
\begin{align}
	\begin{split}
	S_{\text{D3}} &= 2\pi i n_0 \Lambda.
	\end{split}
\end{align}
Solving for $\varphi_{2,\text{BH}}$ and $\varphi_{3,\text{BH}}$ in the equation for $I=2$ and $I=3$, we find
\begin{align}
	\begin{split}
		\varphi_{2,\text{BH}} &= -\omega_{1,\text{BH}} \frac{\Lambda-Q_{3,\text{D3}}}{N}, \qquad \varphi_{3,\text{BH}} = -\omega_{1,\text{BH}} \frac{\Lambda-Q_{2,\text{D3}}}{N}.
	\end{split}
\end{align}
and imposing this into the equation for $k=1$, we have
\begin{align}
	\begin{split}
	0 &= \left(\Lambda - Q_{2,\text{D3}}\right)\left(\Lambda - Q_{3,\text{D3}}\right) + N(J_{1,\text{D3}} + \Lambda)
	\\
	&= \Lambda^2 + \Lambda (-Q_{2,\text{D3}}-Q_{3,\text{D3}} + N) + (Q_{2,\text{D3}}Q_{3,\text{D3}} + N  J_{1,\text{D3}}).
	\end{split}
\end{align}
As this is a quadratic polynomial with real coefficients, the solutions can either be two real roots or two complex roots, conjugate to each other. Therefore, we have
\begin{align} \label{eq:appendixLambdasol}
    \Lambda_{\pm} &= \frac{1}{2} \left(Q_{2,\text{D3}}+Q_{3,\text{D3}}-N \pm \sqrt{\left(-Q_{2,\text{D3}}-Q_{3,\text{D3}}+N\right){}^2 - 4 \left(N J_{1,\text{D3}}+Q_{2,\text{D3}} Q_{3,\text{D3}}\right)}\right).
\end{align}
In general, using the expression for the charges found in \cite{Chen:2023lzq}, $\Lambda_{\pm}$ are complex valued roots, but not necessarily complex conjugate to each other. Moreover, we have the extremization equations
\begin{align} \label{eq:appendixQJsol}
	\Lambda = Q_{1,\text{D3}}\,, \qquad \Lambda = -J_{2,\text{D3}}\,.
\end{align}
In the regime that \eqref{eq:appendixLambdasol} and \eqref{eq:appendixQJsol} are satisfied, we find a complex-valued entropy.

%%%%%%%%%%%%%%%%%%%%%%%%%%%%%%%%
%%%%%%%%%%%%%%%%%%%%%%%%%%%%%%%%%%%%
%%%%%%%%%%%%%%%%%%%%%%%%%%%%%%%%%%%%

\section{Elliptic functions and their asymptotic behavior} \label{App:elliptic:functions}
%%%%%
Here we gather definitions and useful identities of elliptic functions. 

The Pochhammer symbol is defined as
\begin{equation}
	(z;q)_{\infty}=\prod_{k=0}^\infty(1-zq^k).\label{def:pochhammer}
\end{equation}
The elliptic theta functions have the following product forms
\begin{subequations}
\begin{align}
	\theta_0(u;\tau)&=\prod_{k=0}^\infty(1-e^{2\pi i(u+k \tau)})(1-e^{2\pi i(-u+(k+1)\tau)})\label{eq:theta:0}\\
	& = \left(e^{2 \pi i u}; \, e^{2 \pi i \tau}\right)_{\infty} \left(e^{2 \pi i (\tau-u)}; \, e^{2 \pi i \tau}\right)_{\infty}\label{eq:theta:01}\,.
\end{align}
\end{subequations}

Consider an asymptotic expansion in $\tau$  with fixed $0<\arg\tau<\pi$ as given in \cite{Garoufalidis:2018qds}:
\begin{align} \label{Eq:Pochhamer-IdentityMain}
    (z {\rm e}^{ a  \pi \i \tau}; {\rm e}^{2 \pi \i \tau})_\infty &= \exp \left(  \frac{1}{2 \pi \i \tau} \sum_{r = 0}^\infty (-1)^r B_r\left(1 - \frac{a}{2}\right) \frac{(2  \pi \i \tau)^r}{r!} \text{Li}_{2-r}(z)\right)\, .
\end{align}
The elliptic gamma function and the `tilde' elliptic gamma function are defined as
\begin{subequations}
\begin{align}
	\Gamma(z;p,q)&=\prod_{j,k=0}^\infty\fft{1-p^{j+1}q^{k+1}z^{-1}}{1-p^jq^kz},\label{def:gamma}\\
	\widetilde\Gamma(u;\wO,\tau)&=\prod_{j,k=0}^\infty\fft{1-e^{2\pi i[(j+1)\wO+(k+1)\tau-u]}}{1-e^{2\pi i[j\wO+k\tau+u]}}.\label{def:tilde:gamma}
\end{align}
\end{subequations}
To study asymptotic behaviors of elliptic functions, we introduce a $\tau$-modded value of a complex number $u$, namely $\{u\}_\tau$, as
\begin{equation}
	\{u\}_\tau\equiv u-\lfloor\Re u-\cot(\arg\tau)\Im u\rfloor\quad(u\in\mathbb C),\label{tau-modded}
\end{equation}
and define $\{x\}$ such that
\begin{equation}
	\{x\}\equiv x-\lfloor \text{Re} x \rfloor \, ,\label{modded}
\end{equation}
where $\{x\}_{\tau} = \{x\}$ for purely imaginary $\tau$.
\

	\bibliographystyle{JHEP}
	
	\bibliography{BHLocalization.bib}

\end{document}